\documentclass{article}

\usepackage{microtype}
\usepackage{graphicx}
\usepackage{subfigure}
\usepackage{booktabs} % for professional tables

\usepackage{hyperref}

\usepackage[accepted]{icml2025}

\usepackage{amsmath}
\usepackage{amssymb}
\usepackage{mathtools}
\usepackage{amsthm}

\usepackage[capitalize,noabbrev]{cleveref}

\theoremstyle{plain}
\newtheorem{theorem}{Theorem}[section]

\theoremstyle{definition}

\theoremstyle{remark}

\usepackage[textsize=tiny]{todonotes}

\usepackage{amsfonts}

\usepackage{listings}
\usepackage{algorithm}
\usepackage{algorithmic}
\usepackage{multirow}
\newcommand{\sys}{\textsc{NSmark}}

\icmltitlerunning{\sys: Null Space Based Black-box Watermarking Defense Framework for Language Models}

\begin{document}

\twocolumn[
\icmltitle{\sys: Null Space Based Black-box Watermarking \\ Defense Framework for Language Models}

% It is OKAY to include author information, even for blind
% submissions: the style file will automatically remove it for you
% unless you've provided the [accepted] option to the icml2025
% package.

% List of affiliations: The first argument should be a (short)
% identifier you will use later to specify author affiliations
% Academic affiliations should list Department, University, City, Region, Country
% Industry affiliations should list Company, City, Region, Country

% You can specify symbols, otherwise they are numbered in order.
% Ideally, you should not use this facility. Affiliations will be numbered
% in order of appearance and this is the preferred way.
\icmlsetsymbol{equal}{*}

\begin{icmlauthorlist}
\icmlauthor{Haodong Zhao}{1,2}
\icmlauthor{Jinming Hu}{1}
\icmlauthor{Peixuan Li}{1}
\icmlauthor{Fangqi Li}{1}
\icmlauthor{Jinrui Sha}{1}
\icmlauthor{Tianjie Ju}{1}
\icmlauthor{Peixuan Chen}{2}
%\icmlauthor{}{sch}
\icmlauthor{Zhuosheng Zhang}{equal,1}
\icmlauthor{Gongshen Liu}{equal,1}
%\icmlauthor{}{sch}
%\icmlauthor{}{sch}
\end{icmlauthorlist}

\icmlaffiliation{1}{Shanghai Jiao Tong University}
\icmlaffiliation{2}{Tencent}

\icmlcorrespondingauthor{Zhuosheng Zhang}{zhangzs@sjtu.edu.cn}
\icmlcorrespondingauthor{Gongshen Liu}{lgshen@sjtu.edu.cn}

% You may provide any keywords that you
% find helpful for describing your paper; these are used to populate
% the "keywords" metadata in the PDF but will not be shown in the document
\icmlkeywords{Machine Learning, ICML}

\vskip 0.3in
]

% this must go after the closing bracket ] following \twocolumn[ ...

% This command actually creates the footnote in the first column
% listing the affiliations and the copyright notice.
% The command takes one argument, which is text to display at the start of the footnote.
% The \icmlEqualContribution command is standard text for equal contribution.
% Remove it (just {}) if you do not need this facility.

\printAffiliationsAndNotice{}  % leave blank if no need to mention equal contribution
% \printAffiliationsAndNotice{\icmlEqualContribution} % otherwise use the standard text.

\begin{abstract}
Language models (LMs) have emerged as critical intellectual property (IP) assets that necessitate protection. Although various watermarking strategies have been proposed, they remain vulnerable to Linear Functionality Equivalence Attack (LFEA), which can invalidate most existing white-box watermarks without prior knowledge of the watermarking scheme or training data. 
This paper analyzes and extends the attack scenarios of LFEA to the commonly employed black-box settings for LMs by considering Last-Layer outputs (dubbed LL-LFEA). 
We discover that the null space of the output matrix remains invariant against LL-LFEA attacks. 
Based on this finding, we propose \sys, a black-box watermarking scheme that is task-agnostic and capable of resisting LL-LFEA attacks. 
{\sys} consists of three phases: (i) watermark generation using the digital signature of the owner, enhanced by spread spectrum modulation for increased robustness; (ii) watermark embedding through an output mapping extractor that preserves the LM performance while maximizing watermark capacity; (iii) watermark verification, assessed by extraction rate and null space conformity. Extensive experiments on both pre-training and downstream tasks confirm the effectiveness, scalability, reliability, fidelity, and robustness of our approach. 
Code is available at 
\href{https://github.com/dongdongzhaoUP/NSmark}{https://github.com/dongdongzhaoUP/NSmark}.
\end{abstract}

\section{Introduction}
% During past decades, with the rapid development of artificial intelligence, language models (LMs)  have achieved superior performance and been applied in a wide range of fields. 
% During past decades, language models (LMs)  have achieved superior performance and been applied in a wide range of fields. 
% At the same time, training high-performance LMs requires a large amount of data and computing resources, thus LMs can be regarded as valuable intellectual property (IP). With the deployment and application of machine learning as a service (MLaaS) platforms, companies sell well-trained LMs as commodities and release APIs for the public to access. Once sold models are illegally stolen, distributed or resold, the rights of the model owner will be seriously infringed. Therefore, it is necessary to protect IP of LMs.

Over the past few decades, language models (LMs) have achieved exceptional performance and found applications across a wide range of fields \cite{yu2025identifying}.
However, training high-performance LMs requires vast amounts of data and significant computational resources, making these models valuable intellectual property (IP). With the rise of machine learning as a service (MLaaS) platforms, companies sell well-trained LMs as commodities and release APIs for public access. Once these models are illegally stolen, distributed or resold, the rights of the model owners are severely violated. Therefore, protecting the intellectual property of LMs is essential.

\begin{figure}
\vskip 0.2in
\begin{center}
\centerline{\includegraphics[width=\columnwidth]{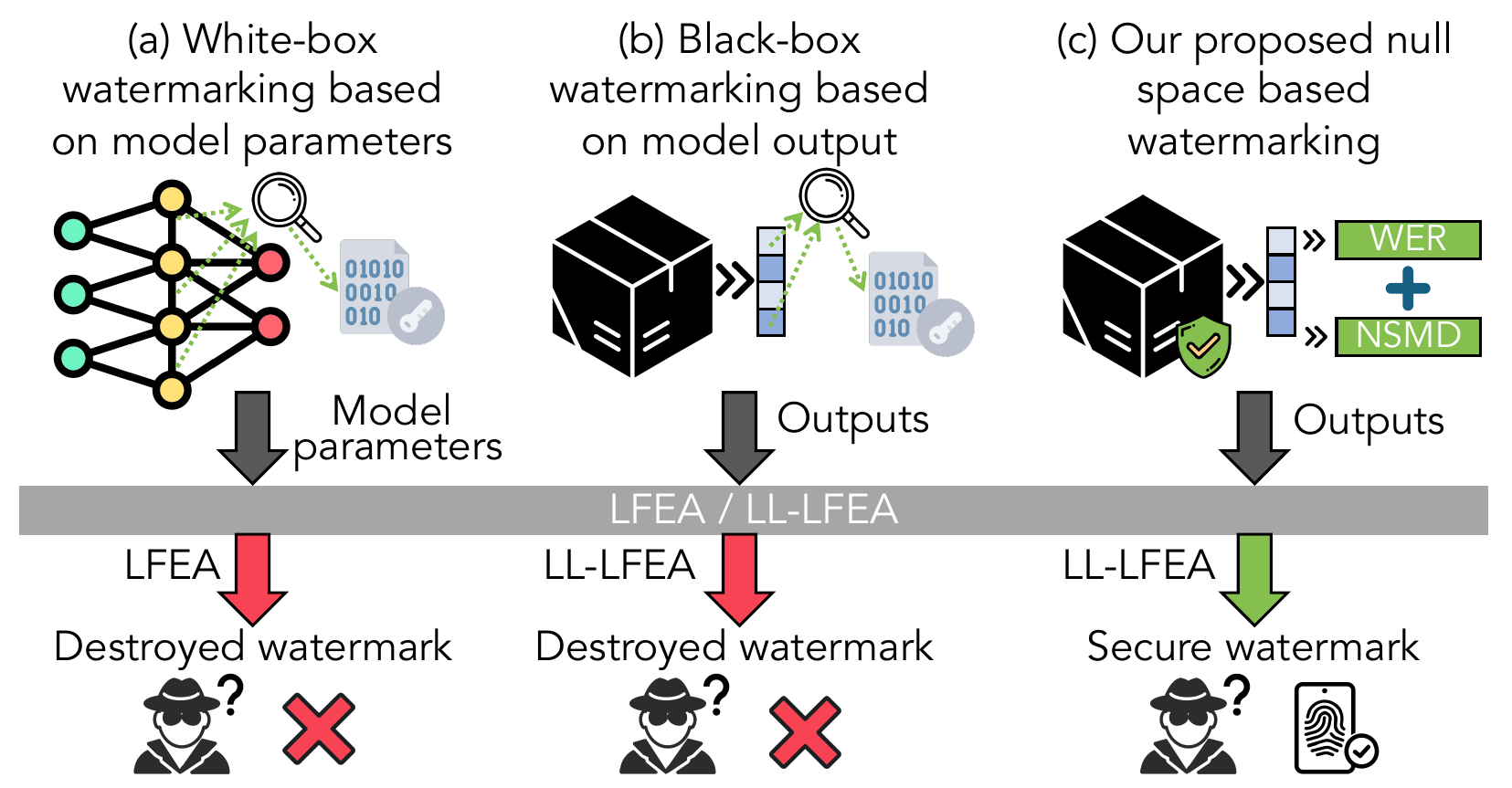}}
\caption{Illustration of different watermark schemes against LFEA/LL-LFEA. LFEA disables parameters based white-box schemes \cite{li2023linear} and LL-LFEA disables output based black-box schemes (Section~\ref{lfea}). \sys\, is secure against LL-LFEA using null space invariance.}
\label{fig:Schematic}
\end{center}
\vskip -0.4in
\end{figure}
Watermarking techniques have been widely used to protect the IP of deep learning models \cite{chen2024deep,he-etal-2024-watermarks,pmlr-v235-carlini24a,pmlr-v235-feng24k}. By incorporating identifiable information, these techniques could verify model ownership and provide proof of authenticity.
Existing watermarking schemes can be categorized into white-box and black-box approaches, depending on whether the model parameters need to be accessed in verification. Among these, black-box schemes are more applicable in real-life scenarios, where model parameters are often inaccessible, such as in cases where models are deployed as APIs.

However, protecting the IP of LMs through watermarking presents significant challenges. 
Since LMs can be deployed for various post-training downstream tasks \cite{zhang2023red}, it is crucial that watermark schemes remain task independent. 
Furthermore, recent studies have revealed vulnerabilities and shortcomings in existing watermarking techniques \cite{li2023linear}. Specifically, the proposed Linear Functionality Equivalence Attack (LFEA) is simple to conduct and can compromise most existing white-box watermarks by exploiting linear invariance without knowledge of the watermarking scheme or the training data. As the hidden states and outputs of the last layer in LM are widely used for classification and generation tasks, we consider them, analyze and expand LFEA scenarios to black-box settings utilizing model outputs (dubbed LL-LFEA).

% \cite{liu2018fine,fan2019rethinking,yang2019effectiveness,shafieinejad2021robustness,chen2021refit,aiken2021neural,liu2021removing,li2023linear,yan2023rethinking}. 
% Firstly, many schemes use white-box solutions and are task-dependent, limiting their applicable scenarios. (2) Most black-box watermarking schemes are ‘zero-bit’ methods. Zero-bit watermark can only detect the presence or absence of the watermark and are easily forged because is does not carry any information. (3) Embedded watermark is easily disturbed and becomes invalid, or has a significant impact on the performance of the original model.
% Firstly, existing schemes commonly rely on white-box solutions and are task-dependent, which restricts their practical applicability. Secondly, the majority of black-box watermarking methods are \textit{zero-bit} schemes. Such zero-bit watermarks merely indicate the presence or absence of a watermark and are prone to forgery since they do not encode additional information. Thirdly, embedded watermarks are susceptible to disturbances that can render them ineffective or significantly degrade the model performance.

In this work, we first explore the characteristics of the model output. We discover that the null space of the matrix composed of the model output vectors is invariant under LL-LFEA. Based on this finding, we propose a new null space verification method that can withstand the LL-LFEA attack. This method uses a new metric, the Null Space Matching Degree (NSMD). NSMD measures the degree of match between the output matrix of the suspicious model and the null space of the protected LM. Finally, we propose \sys, a null-space-based task-agnostic black-box watermarking scheme for LMs. 
\sys\, uses identity information to generate all elements related to the watermark and uses the Watermark
Extracting Rate (WER) and NSMD to verify the watermark, thus can pass through as shown in Figure~\ref{fig:Schematic}.
Spread spectrum modulation technology and an extra extractor are also introduced to enhance watermark performance.

Our contributions are summarized as follows:

(i) We analyze the threat of LFEA on output-based watermark and propose LL-LFEA, which can destroy the watermark embedded in the output vector without affecting the performance of downstream tasks.

(ii) We find that the null space of the matrix composed of the output vectors of the model is invariant under LL-LFEA and thus propose a new null space verification method \sys\, which can resist LL-LFEA. Notably, \sys\, is task-agnostic that uses both new null space verification and signature verification to resist LL-LFEA.

(iii) We conduct comprehensive experiments by applying \sys\, to various models of pre-training and downstream tasks. The experimental results demonstrate the effectiveness, fidelity, reliability, and robustness of \sys.

\section{Related Work}

\textbf{Watermarking for LMs.}
With the rise of pre-training in NLP, recent work has explored watermarking specific to LMs. BadPre \cite{jia2022badencoder} introduced a task-agnostic backdoor attack only for MLM-based LMs.
Hufu \cite{xu2024hufu} introduced a modality-agnostic approach for pre-trained Transformer models using the permutation equivariance property.
Explanation as a Watermark \cite{shao2024explanation} addressed the limitations of backdoor-based techniques by embedding multi-bit watermarks into feature attributions using explainable AI.
\cite{peng2023you,shetty2024wet} proposed Embeddings-as-a-Service (EaaS) watermarks to protect the intellectual property of EaaS providers.
\cite{shen2021backdoor,zhang2023red} proposed task-agnostic backdoor attacks by assigning high-dimensional vectors as trigger set labels, but their effectiveness is sensitive to downstream classifier initialization.
\cite{wang2021riga} introduced an auxiliary neural network for watermark embedding using weights from the main network.
\cite{wu2022watermarking} proposed a task-agnostic embedding loss function, but didn't consider the need for triggers to reflect the model owner's identity. \cite{cong2022sslguard} introduced a black-box watermarking scheme for PLMs, but its applicability is limited due to the discrete nature of word tokens. Unfortunately, these schemes are vulnerable to attacks by LFEA or LL-LFEA in principle.

\noindent\textbf{Watermark Removal Attacks.}
DNN watermarking faces various removal attempts. Common methods include fine-tuning \cite{adi2018turning} and pruning \cite{han2015learning}. Fine-pruning \cite{liu2018fine} combines these approaches for greater effectiveness. Knowledge Distillation \cite{hinton2015distilling} techniques can also inadvertently remove watermarks while reducing model size.
\cite{shetty2024wet} show that existing EaaS watermarks can be removed by paraphrasing attack.
\cite{lukas2022sok} propose a new attack method called Neuron Reordering to swap neurons within the same hidden layer of a DNN to disrupt embedded watermarks in the model's parameters. \cite{li2023linear} introduce a powerful LFEA for white-box watermarks, applying linear transformations to model parameters, effectively destroying embedded watermarks while preserving the model's original functionality. 
Fraud attacks include overwriting \cite{wang2019attacks} and ambiguity attacks \cite{zhu2020secure} also pose a great threat to watermarks.

\begin{figure}
\vskip 0.1in
\begin{center}
\centerline{\includegraphics[width=\columnwidth]{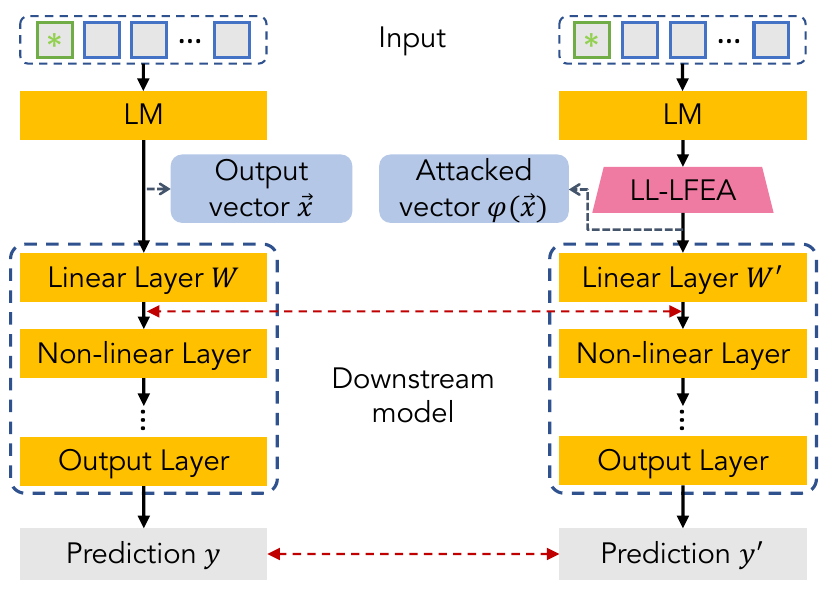}}
\caption{The schematic diagram of model inference flow before and after LL-LFEA attack. LL-LFEA transforms the LM output and performs an inverse transform in the subsequent linear layer, leaving the final prediction unchanged.}
\label{fig:lllfea}
\end{center}
\vskip -0.3in
\end{figure}

\section{Method}
\subsection{Threat Model}
\label{lfea}
In white-box watermarking schemes, high-dimensional model parameters are often used as watermark information. For LMs, since the output of the last layer is also high-dimensional, we can use a method similar to the white-box schemes to embed watermarks in the output. However, embedding identity information into the high-dimensional output vector will face the threat of LFEA-like attacks, which is proposed to destroy watermark information embedded in model parameters by linearly transforming parameters of intermediate layers. Next, we discuss the specific form of linear isomorphism attacks in this scenario.

Assume that the attacker knows that the watermark information is embedded in the LM output and seeks to remove the watermark with minimal attack cost (without modifying the model structure or fine-tuning the model) while ensuring that the model's normal task performance remains unaffected. As shown in Figure~\ref{fig:lllfea}, we propose an attack method that satisfies this requirement and provide a proof below. 

The output vector $\vec{x}$ is generated by the LM and serves as input to the downstream model. 
After passing through a series of linear and non-linear layers, the prediction result $y$ is obtained.
The attacker attempts to modify the output vector of the LM to destroy the watermark while ensuring that the final prediction remains unaffected.
Specifically, the attacker changes $\vec{x}$ to $\varphi(x)$ and inputs it into the downstream model, so that the resulting prediction $y^{\prime}$ remains equal to the original prediction $y$.

The sufficient condition for this result is that the modification to the LM output vector is compensated for after passing through the first linear layer of the downstream network. Let the parameter matrix of the first linear layer in the downstream network be denoted by $W$. In this case, the attacker aims to satisfy the following condition: 
$
W^{\prime}\varphi(\vec{x})=W\vec{x},
$
which leads to 
$
\varphi(\vec{x})=W^{\prime,\dagger}W\vec{x}=Q\vec{x},
$
 where $Q=W^{\prime,\dagger}W$ and $W^{\prime,\dagger}$ is the pseudo-inverse of $W^{\prime}$ \cite{li2023linear}. To avoid loss of information during the linear transformation (since this would adversely affect downstream tasks), $Q$ must be a reversible matrix. We present a simple method and analysis on how to quickly generate high-dimensional Q in Appendix~\ref{appendix-Q}. 

We show that the attacker can apply a linear transformation to $\vec{x}$ thereby destroying the watermark embedded in the output vector, while leaving the downstream task performance unchanged. We refer to this attack as the Last-Layer Linear Functionality Equivalence Attack (LL-LFEA). In addition to the theoretical analysis, the effectiveness of LL-LFEA is experimentally verified in Appendix~\ref{apd-attack}.

\subsection{Null Space Verification Theory}
LL-LFEA applies a linear transformation to the output vector of LM and can destroy the embedded watermark. As a result, previous watermark verification methods may be significantly impacted. We observe that the null space of the matrix composed of the output vector is invariant under the LL-LFEA attack. Based on this, we propose to use the null space matching degree to verify whether the model is embedded with watermarks.
% \begin{thm}
% Before and after LL-LFEA, the null space of the output matrix of LM remains unchanged for the same input set. (Proofs in Appendix~\ref{Null-Space-Verification-Theory})
% \end{thm}

\begin{theorem}
\label{thm:bigtheorem}
Before and after LL-LFEA, the null space of the output matrix of LM remains unchanged for the same input set. 
\end{theorem}
\vspace{-2.5mm}
\begin{proof} 
See details in Appendix~\ref{Null-Space-Verification-Theory}.
\end{proof}
\vspace{-2.5mm}
\begin{figure*}[!ht]
\vskip 0.1in
\begin{center}
\centerline{\includegraphics[width=2.07\columnwidth]{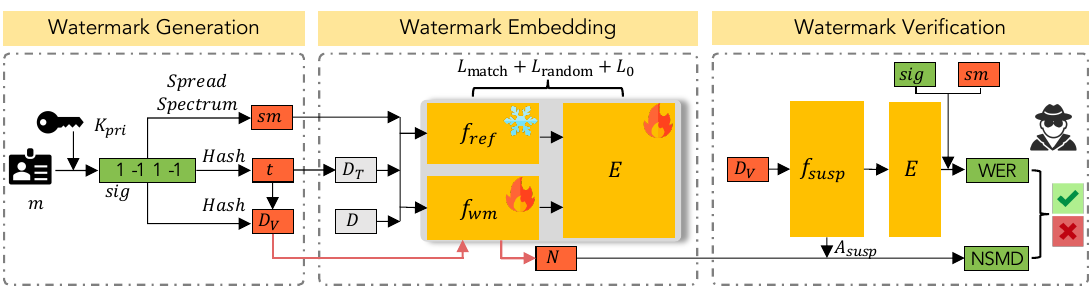}}
\caption{The overall workflow of \sys. (i) In watermark generation, identity information is used generate $sig$. (ii) In watermark embedding, watermarked model $f_{wm}$ and extractor $E$ are trained with the participation of the reference model $f_{ref}$. (iii) In watermark verification, WER and NSMD collaborate to verify the identity of the model.}
\label{fig:workflow}
\end{center}
\vskip -0.3in
\end{figure*}

Therefore, even if the watermark based on the digital string is corrupted, we can still verify model ownership using the null space of the output matrix.
\subsection{Null Space Match Degree (NSMD)}
We define NSMD by introducing the distribution of elements in a matrix, which is obtained by multiplication of the matrix of the output matrix $A$ of any LM without watermark and the null space matrix $N$ of $f_{wm}$. In $H_{(n\times p)}=A_{(n\times m)} \times N_{(m\times p)}$, $H_{i,j}=\alpha_i \cdot \beta_j$ is the dot product of the $i$th row vector of $A$ and the $j$th column vector of $N$. 
We define NSMD of $A$ and $N$ as:
\begin{equation}
\label{eqnsmd}
    % NSMD(A,N)=\frac{1}{n}\sum\limits_{i=1}^{n}\sqrt{\sum\limits_{j=1}^{p} h_{i,j}^2}
    \text{NSMD}(A,N)=\frac{1}{n}\sum\limits_{i=1}^{n}\sum\limits_{j=1}^{p} \sqrt{|H_{i,j}|}.
\end{equation}

Furthermore, we give a detailed analysis of estimation of NSMD (in Appendix~\ref{Estimation-of-NSMD}). For example, if $n=768$ and $p=1500$, we have NSMD $>27.48$. If $N$ is the null space matrix of $A$, NSMD is a minimum value close to $0$. This difference is amplified by the process of calculating the square root, resulting in a significant difference between whether $A$ and $N$ are matched. We use this difference to distinguish whether the model is embedded with a watermark.

\subsection{Overall Framework of \sys}
\sys\, includes three modules: watermark generation, watermark embedding, and watermark verification, as shown in Figure~\ref{fig:workflow}. We describe the modules as follows.
\subsubsection{Watermark Generation}
\label{wmgen}
Algorithm~\ref{alg:gene} shows the watermark generation workflow. We hope that the generated watermark contains the owner's identity information. First, the digital signature $sig = \textbf{Sign}(m)$ is generated from the identity information message $m$. To ensure that the trigger $t$ has a unique mapping relationship with $sig$, only one trigger is used. We use the trigger generation algorithm $\textbf{Encode}(\cdot)$ introduced in \cite{li2023plmmark} to obtain $t = \textbf{Encode}(sig,n = 1)$. $t$ is inserted into clean sample $x$ of dataset $D$ to form a trigger set $D_T$.

To defend against ambiguous attacks, the verification trigger set $D_V$ used for the null space verification also needs to be generated based on $sig$. A candidate pool $D_{NS}$ to generate null space verification data sets should be published, and then a fixed number of samples are selected from the $D_{NS}$ based on the digital signature as the verification data set $D_V$. We define the verification data set selection algorithm as $\textbf{Select}(sig)\rightarrow D_V$, which must be a deterministic algorithm, that is, for the same input, there must be the same output. In addition, we hope that the algorithm will have different outputs for different inputs. Therefore, we choose a hash function and use a one-way hash chain to generate $D_V$. We hope that the index repetition rate obtained by different hash-value mappings is low, so we hope that the data set $D_{NS}$ is as large as possible. The specific process of the $\textbf{Select}(\cdot)$ algorithm is shown in Algorithm~\ref{alg:select}.

\begin{figure}
\vskip 0.1in
\begin{center}
\centerline{\includegraphics[width=0.9\columnwidth]{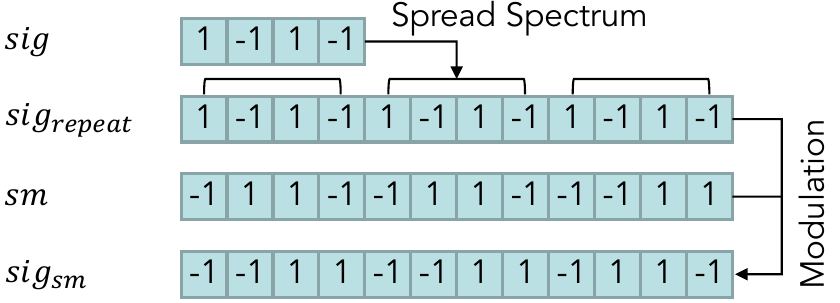}}
\caption{Example diagram of spread spectrum modulation. Repeat $sig$ to obtain $sig_{repeat}$, then use $sm$ to modulate $sig_{repeat}$ to obtain the spread spectrum modulated digital signature $sig_{sm}$.}
\label{fig:ssm}
\end{center}
\vskip -0.3in
\end{figure}

To improve the robustness of the watermark, we introduce the spread spectrum modulation technology as \cite{feng2020watermarking}. Spread spectrum modulation technology uses redundant bits to represent the original information. Figure~\ref{fig:ssm} shows an example of the spreading of $3\times$. 
Please refer to Appendix~\ref{Process-of-Spread-Spectrum-Modulation} for the specific process $\textbf{SM}(sig) \rightarrow sig_{wm}$.

\subsubsection{Watermark Embedding}
Before the training starts, make a copy of $f_{wm}$ as the frozen reference model $f_{ref}$. Then use the clean data set $D$ and the trigger set $D_T$ to train $f_{wm}$ and the extractor $E$. 
When taking $\{D, D_T\}$ as input, $f_{wm}$ will output $\{V, V^T\}$ and $f_{ref}$ will output $\{V_{ref}, V_{ref}^T\}$, respectively. For $V^T$, $E$ maps it to obtain the signature $sig_{wm}$, and for $\{V, V_{ref}, V_{ref}^T\}$, $E$ maps them to random vectors. 
% The output representation of $f_{wm}$ to the trigger set $D_T$ is passed through $E$ to obtain the signature $sig_{wm}$, and $E$ maps the output vector of clean data set $D$ passing $f_{wm}$ and the output representation of $D$ and $D_T$ passing $f_{ref}$ into random vectors.
After the training is completed, $f_{wm}$ is embedded with the watermark. Then using $D_V$ as input, the output vectors are concatenated into a matrix $A$, and the corresponding null space matrix $N$ of $A$ is calculated as part of the key.

Three networks are involved in watermark embedding: the model $f_{wm}$ to be embedded with the watermark, the reference model $f_{ref}$ and the extractor model $E$. Compared to directly embedding $sig$ in the output vector of $f_{wm}$, adding $E$ to the map can reduce the side effect of the watermark on the original performance. The watermark capacity is increased at the same time. 
We use the mean square error loss (MSE) and the similarity function $\operatorname{sim}$ to implement the above training process:
\begin{equation}
\label{lossmatch}
    L_{\text {match}}=\frac{1}{|D_T|} \sum_{x \in D_T} \operatorname{MSE}\left(E\left(V^T\right), {sig}_{s m}\right),
\end{equation}
\begin{equation}
\begin{aligned}
L_{\text {random }} & =\frac{1}{|D|} \sum_{x \in D} \operatorname{sim}\left(E\left(V\right), sig_{sm}\right)^2 \\
& +\frac{1}{|D_T|} \sum_{x \in D_T} \operatorname{sim}\left(E\left(V_{ref}^T\right), sig_{sm}\right)^2 \\
& +\frac{1}{|D|} \sum_{x \in D} \operatorname{sim}\left(E\left(V_{ref}\right), sig_{sm}\right)^2.
\end{aligned}
\end{equation}

We use cosine similarity as the $\operatorname{sim}$ function. The complete loss function of $E$ is $L_{\text{Extractor}}=\lambda_1 L_{\text {match}}+(1-\lambda_1)L_{\text {random}}$. During training, only the parameters of $E$ are trainable. The loss of $f_{wm}$ also consists of two parts: $L_{wm}=\lambda_2 L_{match}+(1-\lambda_2)L_0$. The content of this $L_{match}$ is the same as $L_{match}$ of $E$, but only the parameters of $f_{wm}$ are updated at this time, and $L_0$ is the original LM training loss function. During training, $E$ and $f_{wm}$ are trained alternately.

\subsubsection{Watermark Verification}
To effectively defend attacks, \sys\, uses two metrics together to verify ownership: WER and NSMD. Model owner needs to submit $key=(sig,E,N)$ to the Certification Authority (CA). CA generates $t,D_V,sm$ using $sig$. Input $D_V$ to the suspicious model $f_{susp}$ to get the output vector $A_{susp}$, and pass $A_{susp}$ through $E$ to get the mapped vector $O_{susp}$. Then WER is obtained from the despread spectrum.

WER is defined from comparing bits in $sig$ and $sig^{\prime}$:
% as the proportion of $sig$ and $sig^{\prime}$ having the same value:
\begin{equation}
\label{eq:wr}
    \text{WER}=\frac{1}{n} \sum_{i =0}^{n-1}\left[a_i=a_i^{\prime}\right],
\end{equation}
where $\left[ \cdot \right]$ is the \textit{inverse bracket}, which is $1$ when the expression in the bracket is \textit{True}, otherwise it is $0$.

NSMD is calculated using $A_{susp}$ and $N$ by Equation~\ref{eqnsmd}. We define two thresholds, and whether WER$>T_{W}$ will be first verified. If it fails, whether NSMD$<T_{N}$ will be further considered in the case of LL-LFEA.

\section{Experiments}
\subsection{Experimental Setup}
\label{sec-setup}
\textbf{Datasets.} 
We use WikiText-2 \cite{merity2017pointer} for pre-training and watermark embedding. To evaluate the performance on downstream tasks, we select many text classification datasets: SST-2 and SST-5 \cite{socher2013recursive} for sentiment analysis, Lingspam \cite{sakkis2003memory} for spam detection, OffensEval \cite{zampieri2019semeval} for offensive language identification and AG News \cite{zhang2015character} for news classification.

\begin{table*}[!ht]
\caption{Effectiveness of \sys\, on different LMs. $f_{wm}$ means watermarked model and $f_{clean}$ is not watermarked. WER=$1.00$ indicates that the signature information can be accurately extracted, and NSMD has obvious differentiation between $f_{wm}$ and $f_{clean}$.}
\label{tab:effective}
\vskip 0.1in
\begin{center}
\begin{small}
\begin{sc}
 \setlength{\tabcolsep}{7pt}
\begin{tabular}{@{}c|cccc|cccc@{}}
\toprule
\multirow{2}{*}{Metric} & \multicolumn{4}{c|}{$f_{wm}$} & \multicolumn{4}{c}{$f_{clean}$} \\ \cmidrule(l){2-9} 
 & BERT & RoBERTa & DeBERTa & XLNet & BERT & RoBERTa & DeBERTa & XLNet \\ \midrule
WER & $1.00$ & $1.00$ & $1.00$ & $1.00$ & $0.00$ & $0.00$ & $0.00$ & $0.03$ \\ \midrule
NSMD & $2.94\times 10^{-6}$ & $2.53\times 10^{-6}$ & $2.91\times 10^{-6}$ & $2.90\times 10^{-6}$ & $60.95$ & $61.24$ & $87.88$ & $76.74$ \\ \bottomrule
\end{tabular}%
\end{sc}
\end{small}
\end{center}
\vskip -0.25in
\end{table*}
\noindent\textbf{Models.}
For LMs, we use the base versions of BERT \cite{kenton2019bert}, RoBERTa \cite{liu2019roberta}, DeBERTa \cite{he2020deberta} and XLNet \cite{yang2019xlnet} for main results. Llama-2-7B \cite{touvron2023llama}, GPT-2 \cite{radford2019language} and the large version of BERT and RoBERTa are also used in supplementary experiments. All pretrained weights are from HuggingFace.\footnote{https://huggingface.co/} The extractor network is a three-layer linear network with hidden layers of neurons $2048$ and $1024$. The input dimension matches the output dimension of the LM. The output dimension matches the size of $sig_{sm}$. 

\noindent\textbf{Watermark settings and training details.}
We select a string containing owner information as the message $m$, for example, "BERT is proposed by Google in 2018". The length of $sig$ is $256$ and then spread spectrum by a factor $k = 3$, resulting in a $768$-bit $sig_{sm}$.
SST-2 is used as the candidate pool $D_{NS}$. $q$, the length of $D_V$, is $1500$. The trigger is inserted into random positions for $5$ times in the trigger set.
When performing watermark embedding, $\lambda_1 = 0.5$ and $\lambda_2 = 0.2$ in $L_{Extractor}$ and $L_{wm}$. The batchsize is $4$, and the learning rates for both $f_{wm}$ and $E$ are $10^{-4}$. $f_{wm}$ and $E$ are trained alternately for the $10$ epochs. When fine-tuning downstream tasks, the learning rate is $2 \times 10^{-5}$ and the batchsize is $8$ for $3$ epochs. 

\noindent\textbf{Metrics.}
As mentioned before, two metrics are defined to verify the identity of the model: WER and NSMD. Besides, we adopt accuracy (ACC, in \%) to measure the performance of LM on downstream tasks. 

\subsection{LL-LFEA Attack Evaluation}

We select the effective \textit{word embedding-based watermarking scheme (EmbMarker)} \cite{peng2023you} and \sys\, (without NSMD) as victims to study the effectiveness of LL-LFEA. The attack results on EmbMarker are shown in Appendix~\ref{apd-attack} and the results on \sys\, (without NSMD) are shown in Table~\ref{tab:down}. Though EmbMarker can pass through the attack of RedAlarm \cite{zhang2023red}, after the LL-LFEA attack, all the metrics of EmbMarker are very close to those of the original model without watermark.  At the same time, LL-LFEA has little degradation on original model performance. This fully demonstrates the effectiveness of LL-LFEA on existing watermarking schemes.

\subsection{\sys\, Performance Evaluation}
We analyze the main experimental results about \sys. For more results on computational cost and other more detailed experiments, please refer to Appendix~\ref{appendix-results}.
\subsubsection{Effectiveness}
Effectiveness means that the watermark can achieve the expected effect during verification. Ideally, $sig^\prime$ extracted from the watermarked model should be consistent with the original $sig$, and the output matrix of $f_{wm}$ for $D_V$ should match completely $N$ stored in $key$, which means WER = $1$ and NSMD = $0$. Table~\ref{tab:effective} shows the results of $f_{wm}$ embedded with watermark and $f_{clean}$ without watermark. It can be seen that for different watermarked LMs, WER is $1$ and NSMD is close to $0$. This shows the effectiveness of \sys. Comparison of the values of $f_{wm}$ and $f_{clean}$ shows that WER and NSMD will obviously change after the watermark is embedded. Although NSMD of different LMs has different values, they are all far from $0$. Through these results, we can preliminarily define verification thresholds of WER and NSMD as $T_{W} = 0.6$ and $T_{N} = 43$, which are $0.6 \times$ the average gaps. Thresholds can be further adjusted according to different models and task types. We also study the watermark effectiveness on larger size of the models in Table~\ref{tab:large}, which demonstrates the scalability of \sys.

\subsubsection{Reliability}
The watermark key is a triple $key = (sig, E, N)$. Next, we analyze whether the watermark can be successfully verified if an attacker provides an incorrect $key$.

\begin{figure}
\vskip 0.2in
\begin{center}
\centerline{\includegraphics[width=1.0\columnwidth]{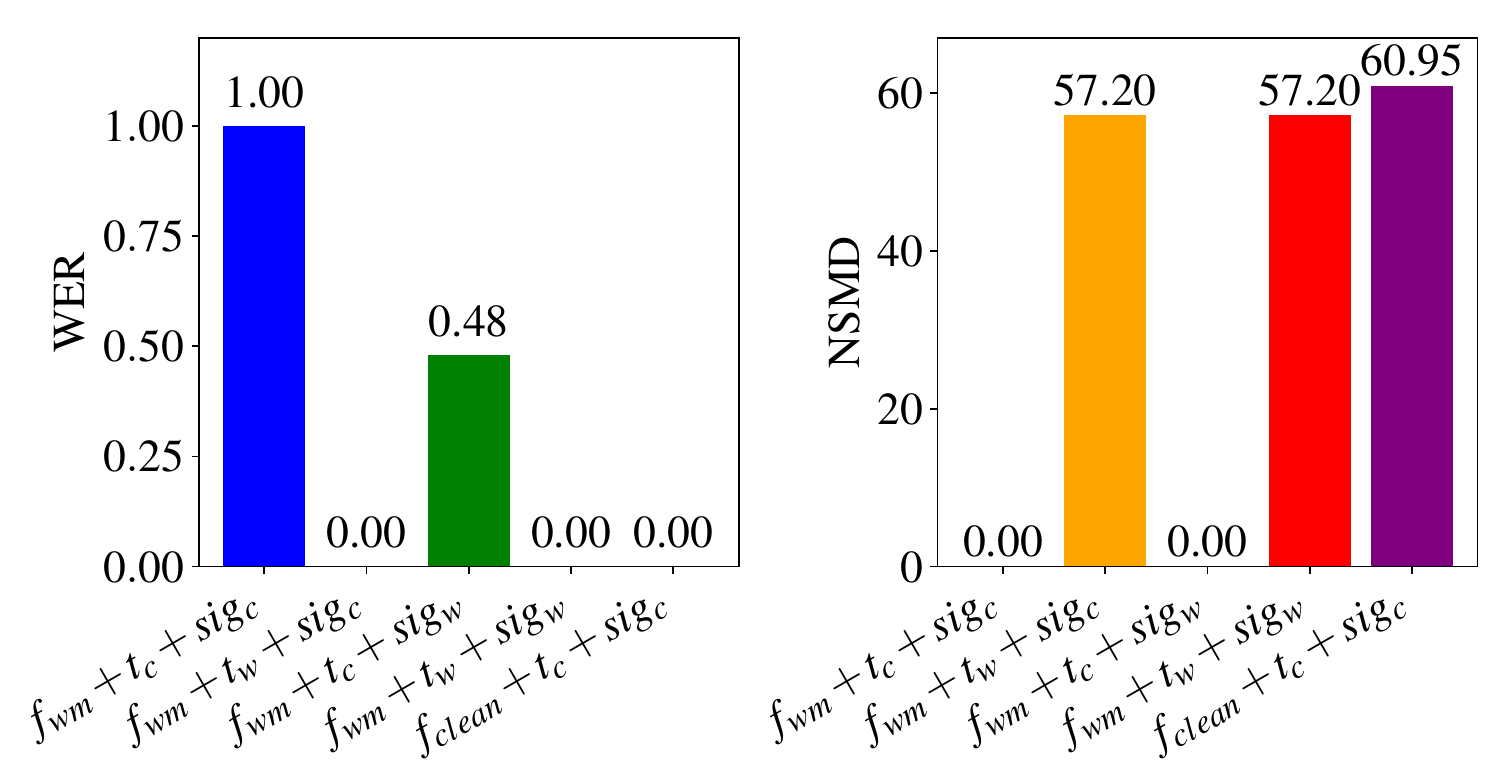}}
\caption{The impact of the correctness of trigger $t$ and signature $sig$ on WER and NSMD. The $c$ in the subscript stands for \textit{correct} and $w$ stands for \textit{wrong}. Only $f_{wm}$ with correct trigger and $sig$ can pass through verification.}
\label{fig:sig}
\end{center}
\vskip -0.35in
\end{figure}
\noindent\textbf{Wrong signature $sig$.}
The trigger $t$ and the output of $f_{wm}$ are related to $sig$, but as $sig$ and $t$ are not a one-to-one mapping relationship, there are situations where only one of $sig$ and $t$ is correct. Figure~\ref{fig:sig} shows all possible scenarios.

For $f_{wm}$, (i) when the trigger is wrong ($t_w$) and the signature is correct ($sig_c$), WER = 0, NSMD $>T_N$. This means that $f_{wm}$ has learned the relationship between $sig$ and $t$. Whether $t$ is correct determines whether $f_{wm}$ can produce the expected output, which in turn affects both the calculation of WER and NSMD.
(ii) When the trigger is correct ($t_c$) and the signature is wrong (considering the most dangerous scenario that $sig_w$ consists only of $\{-1,1\}$), WER $\approx 0.5$. This is because $t_c$ leads to the right $sig^\prime$, and its expectation of WER with a random string $\{-1,1\}$ is $0.5$. As the output matrix is correctly generated by $f_{wm}$ based on $t_c$, NSMD = $0$ in this case. 
(iii) When both trigger and signature are wrong ($t_w$ and $sig_w$), WER = $0$, NSMD$>T_N$, indicating that the watermark cannot be correctly verified without providing the correct key. (iv) For a model without embedded watermarks $f_{clean}$, watermarks cannot be extracted even if the correct key is provided.

\begin{table}[]
\caption{WER results of different extractor $E$. $f_{wm}$, $E_c$, and $E_w$ means watermarked model, correct $E$ and wrong $E$, respectively. Only the correct E can accurately extract the signature.}
\label{tab:e}
\vskip 0.1in
\begin{center}
\begin{small}
\begin{sc}
\resizebox{\columnwidth}{!}
{%
 \setlength{\tabcolsep}{10pt}
\begin{tabular}{@{}ccccc@{}}
\toprule
Setting & BERT & RoBERTa & DeBERTa & XLNet \\ \midrule
$f_{wm}+E_c$ & 1.00 & 1.00 & 1.00 & 1.00 \\
$f_{wm}+E_w$ & 0.00 & 0.00 & 0.00 & 0.13 \\ \bottomrule
\end{tabular}%
}
\end{sc}
\end{small}
\end{center}
\vskip -0.1in
\end{table}

\noindent\textbf{Wrong extractor $E$.}
Since $E$ is not involved in NSMD calculation, we only analyze the impact of $E$ on WER. As shown in Table~\ref{tab:e}, when $E$ is wrong, the WER is close to $0$, indicating that wrong $E$ is unable to extract the watermark.

\begin{table}[]
\caption{NSMD of different null space matrix $N$. $f_{wm}$, $N_c$, $N_r$ and $N_s$ means watermarked model, correct $N$, random $N$ and $N$ composed of small elements, respectively. Only when the correct $N$ is used can the NSMD be close to $0$.}
\label{tab:N}
% \vskip 0.1in
\begin{center}
\begin{small}
\begin{sc}
\resizebox{\columnwidth}{!}{%
\begin{tabular}{@{}ccccc@{}}
\toprule
Setting & BERT & RoBERTa & DeBERTa & XLNet \\ \midrule
$f_{wm}+N_c$ & $2.94\times 10^{-6}$ & $2.53\times 10^{-6}$ & $2.91\times 10^{-6}$ & $2.90\times 10^{-6}$ \\
$f_{wm}+N_r$ & 3167.81 & 3171.58 & 3182.79 & 3117.61 \\
$f_{wm}+N_s$ & 1001.75 & 1002.94 & 1006.49 & 985.87 \\ \bottomrule
\end{tabular}%
}
\end{sc}
\end{small}
\end{center}
\vskip -0.3in
\end{table}

\begin{table}[]
\caption{Impact of fine-tuning on watermark performance. $F_{wm}$ means fine-tuned whole watermarked model and $F_{clean}$ denotes fine-tuned model without watermark. (i) $F_{wm}$ has a slight loss in ACC compared to $F_{clean}$. (ii) Fine-tuning has little effect on the WER of $F_{wm}$. (3) Fine-tuning increases the NSMD, but it is still significantly different from the model without watermark.}
\label{tab:down}
% \vskip 0.1in
\begin{center}
\begin{small}
\begin{sc}
\resizebox{\columnwidth}{!}{%
\begin{tabular}{@{}c|cc|ccccc@{}}
\toprule
Metric & \multicolumn{2}{c|}{Model Setting} & SST-2 & SST-5 & Offenseval & Lingspam & AGnews \\ \midrule
\multirow{8}{*}{ACC} & \multicolumn{1}{c}{\multirow{2}{*}{BERT}} & $F_{wm}$ & 91.40 & 52.62 & 85.12 & 99.14 & 93.95 \\ 
 & \multicolumn{1}{c}{} & $F_{clean}$ & 91.63 & 53.03 & 84.07 & 99.66 & 94.38 \\ \cmidrule(l){2-8}
 & \multicolumn{1}{c}{\multirow{2}{*}{RoBERTa}} & $F_{wm}$ & 92.55 & 54.71 & 84.30 & 99.66 & 94.43 \\  
 & \multicolumn{1}{c}{} & $F_{clean}$ & 94.04 & 56.15 & 84.88 & 100.00 & 94.72 \\ \cmidrule(l){2-8}
 & \multicolumn{1}{c}{\multirow{2}{*}{DeBERTa}} & $F_{wm}$ & 93.00 & 55.48 & 83.02 & 99.31 & 94.61 \\  
 & \multicolumn{1}{c}{} & $F_{clean}$ & 93.58 & 57.65 & 85.12 & 99.31 & 94.84 \\ \cmidrule(l){2-8}
 & \multicolumn{1}{c}{\multirow{2}{*}{XLNet}} & $F_{wm}$ & 88.65 & 42.67 & 81.98 & 99.14 & 93.29 \\ 
 & \multicolumn{1}{c}{} & $F_{clean}$ & 93.58 & 53.62 & 84.65 & 99.31 & 94.07 \\ \midrule
 \multirow{8}{*}{WER} & \multicolumn{1}{c}{\multirow{2}{*}{BERT}} & $F_{wm}$ & 1.00 & 1.00 & 1.00 & 0.94 & 1.00 \\ 
 & \multicolumn{1}{c}{} & $F_{clean}$ & 0.00 & 0.00 & 0.00 & 0.00 & 0.00 \\ \cmidrule(l){2-8}
 & \multicolumn{1}{c}{\multirow{2}{*}{RoBERTa}} & $F_{wm}$ & 0.98 & 1.00 & 1.00 & {0.72} & 0.99 \\ 
 & \multicolumn{1}{c}{} & $F_{clean}$ & 0.00 & 0.00 & 0.00 & 0.00 & 0.00 \\ \cmidrule(l){2-8}
 & \multicolumn{1}{c}{\multirow{2}{*}{DeBERTa}} & $F_{wm}$ & 1.00 & 1.00 & 1.00 & 0.80 & 0.88 \\  
 & \multicolumn{1}{c}{} & $F_{clean}$ & 0.00 & 0.00 & 0.00 & 0.00 & 0.00 \\ \cmidrule(l){2-8}
 & \multicolumn{1}{c}{\multirow{2}{*}{XLNet}} & $F_{wm}$ & 1.00 & 1.00 & 1.00 & 0.92 & 1.00 \\ 
 & \multicolumn{1}{c}{} & $F_{clean}$ & 0.00 & 0.00 & 0.00 & 0.01 & 0.00 \\ \midrule
\multirow{8}{*}{NSMD} & \multicolumn{1}{c}{\multirow{2}{*}{BERT}} & $F_{wm}$ & 29.77 & 25.29 & 22.52 & 21.96 & 24.37 \\ 
 & \multicolumn{1}{c}{} & $F_{clean}$ & 72.97 & 70.06 & 66.65 & 69.90 & 61.59 \\ \cmidrule(l){2-8}
 & \multicolumn{1}{c}{\multirow{2}{*}{RoBERTa}} & $F_{wm}$ & {50.17} & 30.97 & 25.15 & 26.78 & {28.43} \\ 
 & \multicolumn{1}{c}{} & $F_{clean}$ & 74.75 & 74.48 & 69.90 & 65.06 & 75.54 \\ \cmidrule(l){2-8}
 & \multicolumn{1}{c}{\multirow{2}{*}{DeBERTa}} & $F_{wm}$ & {31.89} & 25.74 & 23.52 & 27.23 & {37.68} \\ 
 & \multicolumn{1}{c}{} & $F_{clean}$ & 80.81 & 76.72 & 72.41 & 68.49 & 76.33 \\ \cmidrule(l){2-8}
 & \multicolumn{1}{c}{\multirow{2}{*}{XLNet}} & $F_{wm}$ & 24.12 & 23.52 & 25.29 & 24.06 & 26.20 \\ 
 & \multicolumn{1}{c}{} & $F_{clean}$ & 76.30 & 74.75 & 69.84 & 78.09 & 74.97 \\ \bottomrule
\end{tabular}
}
\end{sc}
\end{small}
\end{center}
\vskip -0.4in
\end{table}

\begin{table*}[]
\caption{LL-LFEA results on \sys. $f_{wm}$ means watermarked model and $f_{LL-LFEA}$ denotes $f_{wm}$ attacked by LL-LFEA. Due to the invariance of the null space to linear transformations, NSMD can still effectively prove the IP of the model despite the failure of WER.}
\label{tab:lfea}
% \vskip 0.1in
\begin{center}
\begin{small}
\begin{sc}
\resizebox{2\columnwidth}{!}{%
\begin{tabular}{@{}c|cc|cc|cc|cc@{}}
\toprule
\multirow{2}{*}{Metric} & \multicolumn{2}{c|}{BERT} & \multicolumn{2}{c|}{RoBERTa} & \multicolumn{2}{c|}{DeBERTa} & \multicolumn{2}{c}{XLNet} \\
 & $f_{wm}$ & $f_{LL-LFEA}$ & $f_{wm}$ & $f_{LL-LFEA}$ & $f_{wm}$ & $f_{LL-LFEA}$ & $f_{wm}$ & $f_{LL-LFEA}$ \\ \midrule
WER & 1.00 & 0.27 & 1.00 & 0.07 & 1.00 & 0.27 & 1.00 & 0.00 \\
NSMD & $2.94\times 10^{-6}$ & 0.06 & $2.53\times 10^{-6}$ & 0.04 & $2.91\times 10^{-6}$ & 0.07 & $2.90\times 10^{-6}$ & 0.05 \\ \bottomrule
\end{tabular}
}
\end{sc}
\end{small}
\end{center}
\vskip -0.3in

\end{table*}
\noindent\textbf{Wrong null space $N$.}
Since $N$ is not involved in calculating WER, we only analyze its impact on NSMD. As shown in Table~\ref{tab:N}, $N_{r}$ is a randomly generated matrix with the same dimension as $N$ and each element is distributed between $[0,1]$. It can be seen that NSMD is very large at this time. However, if the attacker knows the watermark algorithm and want to reduce NSMD, a $N_s$ with extremely small elements might be generated. In this case, NSMD might meet the verification requirements. This indicates that NSMD cannot be used independently to verify the watermark.
\subsubsection{Fidelity}
\label{sec:fidelity}

We hope that \sys\, does not affect the performance on the original tasks. Thus, we add a downstream network to $f_{wm}$, and fine-tune the whole model $F_{wm}$ with the downstream dataset. $F_{clean}$ without watermark is fine-tuned as baseline. Table~\ref{tab:down} shows that the watermark has almost no impact on the performance of the model on the original task. 

\subsubsection{Defense against LL-LFEA}
\noindent\textbf{Defense against LL-LFEA.}
When designing \sys, we focus on resisting LL-LFEA and propose null space verification using NSMD. Table~\ref{tab:lfea} shows the impact of LL-LFEA on watermark verification, where $f_{LL-LFEA}$ denotes the model $f_{wm}$ attacked by LL-LFEA. Experiments show that after LL-LFEA, WER drops significantly, as discussed in Section~\ref{lfea}, but NSMD is still close to $0$, verifying that NSMD is an effective indicator for LL-LFEA. Furthermore, after applying LL-LFEA, the attacker can add a network to $f_{LL-LFEA}$ and fine-tune it for downstream tasks (detailed results are presented in Appendix~\ref{LL-LFEA+finetuning}). Additionally, since LL-LFEA causes minimal degradation in model performance, the attacker may attempt to further compromise the watermark through multiple LL-LFEA attacks. Analysis for this aspect is provided in Appendix~\ref{appendix-multi}.

\begin{figure}
\vskip 0.1in
\begin{center}
\centerline{\includegraphics[width=1.0\columnwidth]{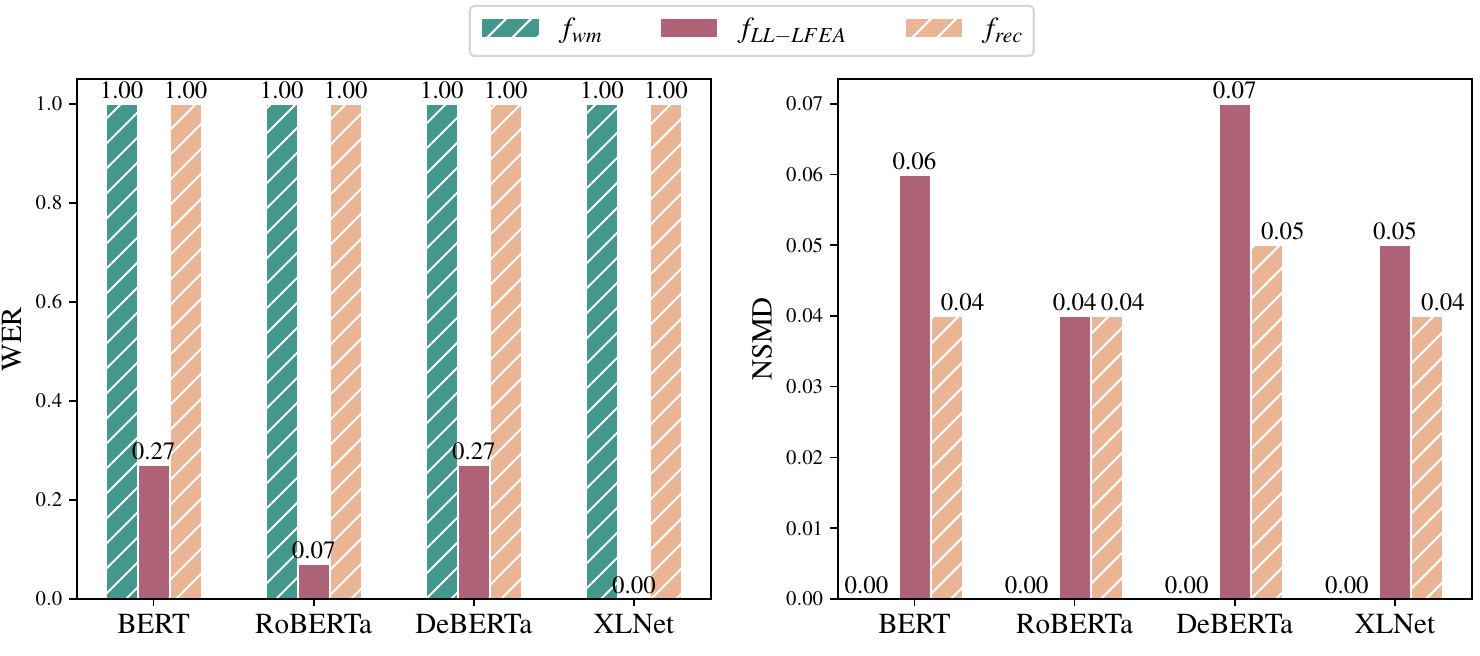}}
\caption{Changes of WER and NSMD before and after LL-LFEA attack and recovery. $f_{wm}$, $f_{LL-LFEA}$ and $f_{rec}$ denote watermarked model, $f_{wm}$ attacked by LL-LFEA, recovered model from $f_{LL-LFEA}$, respectively.}
\label{fig:rec}
\end{center}
\vskip -0.4in
\end{figure}

\noindent\textbf{Recovery of WER.}
In LFEA \cite{li2023linear}, a method is proposed to recover the watermark. We revise this method to recover $f_{rec}$ from $f_{LL-LFEA}$. Specifically, assume that the output matrix of $f_{wm}$ is $A_{1(n\times m)}$ as Proof~\ref{proof}. After being attacked with $Q_{(n\times n)}$, the output matrix turns to $A_{2} = Q \times A_{1}$. Therefore, an estimate of $Q$ can be obtained as $Q^\prime = A_2 \times A_1^{-1}$. If $m \neq n$, then $A_1$ is not reversible and $Q^\prime = A_2 \times A_1^T \times (A_1 \times A_1^T )^{-1}$. Then we perform an anti-attack transformation on $f_{LL-LFEA}$, that is, multiply all the outputs of $f_{LL-LFEA}$ by $Q^\prime$ to get $f_{rec}$. Figure~\ref{fig:rec} shows that after recovery, WER is significantly improved, indicating that such linear attacks are recoverable. In all cases, the NSMD is quite small, proving that NSMD is invariant to LL-LFEA. In the recovery algorithm in \cite{li2023linear}, both the attacker and the model owner might use such an algorithm to claim to be the owner of the model, which will cause verification ambiguity. However, our proposed NSMD is invariant under LL-LFEA, so as long as the timestamp information is added to the $key$ tuple, the ownership can be reliably verified according to the time sequence of the model and the release of $key$.

\subsubsection{Robustness}
The robustness of watermark refers to whether watermark can be effectively verified after watermark removal attacks. Next, we will analyze the robustness of \sys\, against fine-tuning, pruning, fine-pruning, and overwriting attacks. More robustness analysis against paraphrasing attack and multi-time LL-LFEA attack are shown in Appendix~\ref{appendix-paraphrasing}-~\ref{appendix-multi}.

\noindent\textbf{Robustness against fine-tuning.}
Table~\ref{tab:down} shows the WER and NSMD results after fine-tuning on downstream tasks. $F_{wm}$ and $F_{clean}$ are obtained the same as in Section~\ref{sec:fidelity}. In most cases, the WER is still very high, indicating that the embedded $sig$ can still be effectively extracted after downstream fine-tuning. However, the WER of RoBERTa on Lingspam task is relatively low. In main results we set the max input length to 128, which is quite shorter than the average length of Lingspam samples (average length of 695.26). Thus it is not sure the model's input includes triggers (possibly truncated). We perform further experiments on increasing the max length of input to 512, and modify the position of trigger to the front. WER increases to more than $0.84$ and $0.87$ respectively. Besides, compared to $F_{clean}$, there is still obvious discrimination. Therefore, for complex tasks, the verification threshold $T_W$ can be slightly lowered.

\begin{figure}
% \vskip 0.1in
\begin{center}
\centerline{\includegraphics[width=0.7\columnwidth]{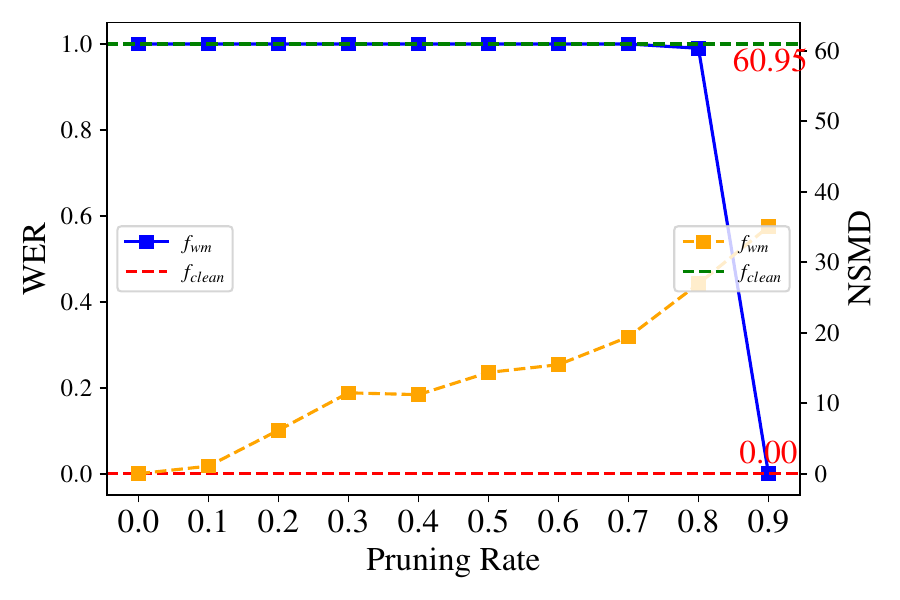}}
\caption{Impact of pruning attacks on watermark. The dotted line is the performance of the original model without watermark.}
\label{fig:prun}
\end{center}
\vskip -0.4in
\end{figure}

\noindent\textbf{Robustness against pruning and fine-pruning.}
Pruning is a commonly used model compression method and is often used to destroy the watermark embedded in the model. Referring to \cite{han2015learning,shao2024explanation}, we sort the parameters of each layer in LM, then set some fractions of parameters with smaller absolute value to $0$. Figure~\ref{fig:prun} shows that when the pruning rate is less than or equal to $0.8$, the WER is close to $1.0$. When the pruning rate is less than or equal to $0.6$, NSMD does not change significantly, and even when the pruning rate is as high as $0.9$, NSMD is still distinguishable. Besides, as shown in Appendix~\ref{appendix-prun}, the accuracy of the watermarked model only changes slightly after pruning then fine-tuning on the SST-5. This shows that the embedded watermark is robust to pruning attack.

Usually, pruning will affect the performance of the model on the original task, and the original task accuracy will be restored by fine-tuning (fine-pruning), as demonstrated by the results  in Appendix~\ref{fine-pruning}.

\noindent\textbf{Robustness against overwriting.}
Overwriting means attacker embeds his own watermark into a model that has already been watermarked in the same way. This may destroy the original watermark. We simulate this process to obtain $f_{ow}$, then add a downstream network and fine-tune to obtain $F_{ow}$. We test the original watermark as shown in Table~\ref{tab:ow}. The overwriting attack has little effect on ACC and WER except on Lingspam. Meanwhile, it has an impact on NSMD similar to that of fine-tuning.

\begin{table}[]
\caption{Impact of overwriting attacks on watermark performance. $f_{ow}$ is the overwritten model, which is fine-tuned to obtain $F_{ow}$. ``--'' means not applicable. }
\label{tab:ow}
\vskip 0.1in
\begin{center}
\begin{small}
\begin{sc}
\resizebox{\columnwidth}{!}
{%
 \setlength{\tabcolsep}{10pt}
\begin{tabular}{@{}ccccc@{}}
\toprule
Model & Downstream dataset & ACC & WER & NSMD \\ \midrule
$f_{ow}$ & -- & -- & 1.00 & 22.76 \\ \midrule
\multirow{5}{*}{$F_{ow}$} & SST-2 & 92.32 & 0.98 & 48.77 \\
 & SST-5 & 50.54 & 1.00 & 36.47 \\
 & Offenseval & 84.77 & 1.00 & 36.61 \\
 & Lingspam & 99.31 & 0.62 & 28.87 \\
 & AGnews & 93.51 & 1.00 & 32.49 \\ \bottomrule
\end{tabular}%
}
\end{sc}
\end{small}
\end{center}
\vskip -0.2in
\end{table}

\section{Further Analysis}
\begin{table}[]
\caption{The necessity of using trigger set in verification. $\text{NSMD}_t$ is the result for trigger set and $\text{NSMD}_c$ is for clean set.}
\label{tab:dv}
\vskip 0.1in
\begin{center}
\begin{small}
\begin{sc}
\resizebox{\columnwidth}{!}{%
\begin{tabular}{@{}c|c|c|ccccc@{}}
\toprule
\multirow{2}{*}{Model} & \multirow{2}{*}{Metric} & \multirow{2}{*}{$f_{wm}$} & \multicolumn{5}{c}{$F_{wm}$} \\ \cmidrule(l){4-8} 
 &  &  & SST-2 & SST-5 & Offenseval & Lingspam & AGnews \\ \midrule
\multirow{2}{*}{BERT} & $\text{NSMD}_t$ & $2.94 \times 10^{-6}$  & 29.77 & 25.29 & 22.52 & 21.96 & 24.37 \\
 & $\text{NSMD}_c$ & $3.01\times 10^{-6}$ & 76.20 & 73.39 & 64.55 & 66.60 & 74.75 \\ \midrule
\multirow{2}{*}{RoBERTa} & $\text{NSMD}_t$ & $2.53\times 10^{-6}$ & 50.17 & 30.97 & 25.15 & 26.78 & 28.43 \\
 & $\text{NSMD}_c$ & $2.54\times 10^{-6}$ & 74.37 & 71.75 & 64.50 & 65.90 & 74.95 \\ \midrule
\multirow{2}{*}{DeBERTa} & $\text{NSMD}_t$ & $2.91\times 10^{-6}$ & 31.89 & 25.74 & 23.52 & 27.23 & 37.69 \\
 & $\text{NSMD}_c$ & $2.98\times 10^{-6}$ & 74.31 & 71.83 & 64.48 & 50.73 & 73.67 \\ \midrule
\multirow{2}{*}{XLNet} & $\text{NSMD}_t$ & $2.90\times 10^{-6}$ & 24.12 & 23.52 & 25.29 & 24.06 & 26.20 \\
 & $\text{NSMD}_c$ & $3.00\times 10^{-6}$ & 68.86 & 55.43 & 41.23 & {24.88} & 38.02 \\ \bottomrule
\end{tabular}%
}
\end{sc}
\end{small}
\end{center}
\vskip -0.2in
\end{table}
Next we discuss the necessity of using trigger set in verification rather than clean set. As shown in Table~\ref{tab:dv}, before downstream fine-tuning $(f_{wm})$,  NSMD for trigger set ($\text{NSMD}_t$) and clean set ($\text{NSMD}_c$) are all close to $0$.After downstream fine-tuning, $\text{NSMD}_c$ is significantly higher than $\text{NSMD}_t$. 

Combining these results, we think fine-tuning has little effect on the output representation of trigger set. However, the output representation of the clean set will change significantly for better performance on different downstream tasks, which causes the null space matrix no longer match the original $N$. Thus $\text{NSMD}_c$ has significantly increase. Therefore, the trigger set is needed for verifying null space.
\section{Conclusion}
This paper proposes \sys, a black-box watermark framework for verification of ownership using the output of LMs. We first analyze and introduce LL-LFEA, and propose a solution that can use null space invariance for watermark verification. We conduct an overall design from three aspects: watermark generation, watermark embedding, and watermark verification. Two indicators, WER and NSMD, are used to jointly verify the existence and identity of the watermark.
Experiments demonstrate the effectiveness, scalability, reliability, and fidelity of \sys, and it has satisfactory performance under various attacks. With the cooperation of two verification methods, a robust and secure watermarking scheme works.

\section*{Impact Statement}
% As discussed in prior works, 
LM watermarking plays a crucial role in IP protection, with its significant societal implications. 
% This paper highlights LL-LFEA threats to existing watermarking schemes, revealing that LM watermarks especially black-box LM watermarks are very needed to be researched and deployed in practice. 
This paper highlights the vulnerabilities of existing watermarking schemes, particularly in the context of LL-LFEA threats, emphasizing the urgent need for further research and practical deployment of robust black-box LM watermarks. 
We believe there is no direct negative impact of making our findings public, nor is there a clear avenue for responsible disclosure. Conversely, we assert that our work contributes positively to society by exposing the vulnerabilities in current watermarking schemes, underscoring the necessity for more resilient designs and rigorous evaluation methodologies. This research represents a significant step toward the practical implementation of robust LM watermarks.

% references without citing it in the main text, use \nocite
% \nocite{langley00}

\bibliography{example_paper}
\bibliographystyle{icml2025}

%%%%%%%%%%%%%%%%%%%%%%%%%%%%%%%%%%%%%%%%%%%%%%%%%%%%%%%%%%%%%%%%%%%%%%%%%%%%%%%
%%%%%%%%%%%%%%%%%%%%%%%%%%%%%%%%%%%%%%%%%%%%%%%%%%%%%%%%%%%%%%%%%%%%%%%%%%%%%%%
% APPENDIX
%%%%%%%%%%%%%%%%%%%%%%%%%%%%%%%%%%%%%%%%%%%%%%%%%%%%%%%%%%%%%%%%%%%%%%%%%%%%%%%
%%%%%%%%%%%%%%%%%%%%%%%%%%%%%%%%%%%%%%%%%%%%%%%%%%%%%%%%%%%%%%%%%%%%%%%%%%%%%%%
\newpage
\appendix
\onecolumn
\section{Additional Details of Theory and Algorithm}
\subsection{The generation of $Q$}
\label{appendix-Q}
In principle, $Q$ needs to be a reversible matrix. To make the value of $Q$ more stable for tensor calculations, we choose to constrain each element to be uniformly distributed between $[0,1]$. The method to generate such a $Q$ is very simple, just sample uniformly between $[0,1]$, and the resulting matrix is a reversible matrix (probability very close to 100\%). The probability that the random matrix $Q$ is singular is the same as a point in $R^{n^2}$ lands in the zero set of a polynomial, which has Lebesgue measure $0$ \cite{fleming2012functions}.
\subsection{Proofs of Null Space Verification Theory}
\label{Null-Space-Verification-Theory}
\begin{proof}
\label{proof}
The null space $N(A)$ of the matrix $A_{(a\times b)}$ is the set of all $b$-dimensional vectors $x$ that satisfy $Ax = \vec{0}$ \cite{axler2015linear}. That is, $N(A) = \{x \in \mathbb{R}^b,Ax = \vec{0}\}$. Using $f_{wm}$ to denote the LM embedded with the watermark, assuming $A_{1(n\times m)}=\{f_{wm}(x),x\in D_T\}$ is the matrix concatenated from the output vectors, where $D_T$ is the verification dataset with watermark trigger, $m$ is the size of $D_T$ and $n$ is the dimension of the output vector of the last layer of the LM. Let the null space matrix of $A_1$ be $N_1$, then $A_1 \times N_1 = \textbf{0}$. 
% Since the matrix of the output of the pre-trained model for the sample is usually a full-rank matrix, for the matrix A with m > n, the dimension of its null space matrix is n× p, where p = m−n.

After performing LL-LFEA, assuming that the new output matrix of $f_{wm}(D_T)$ is $A_{2(n\times m)}$, according to Section~\ref{lfea}, we have $A_2=Q\times A_1$. Then $A_2 \times N_1=(Q \times A_1)\times N_1=Q\times (A_1 \times N_1)=\textbf{0}$, which means that $N_1$ belongs to the null space matrix of $A_2$. As $Q$ is a reversible matrix, then $rank(A_1)=rank(A_2)$, and the null spaces of $A_1$ and $A_2$ have the same dimension. It can be concluded that $N_1$ is also the null space matrix of $A_2$. 
\end{proof}

\subsection{Estimation of NSMD}
\label{Estimation-of-NSMD}
We define NSMD by introducing the distribution of elements in a matrix, which is obtained by matrix multiplication of the output matrix $A$ of any LM without watermark and the null space matrix $N$ of $f_{wm}$. In $H_{(n\times p)}=A_{(n\times m)} \times N_{(m\times p)}$, $H_{j,j}=\alpha_i \cdot \beta_j$ is the dot product of the i-th row vector of $A$ and the j-th column vector of $N$. 
The approximate distribution of the angle between $n$ random uniformly distributed unit vectors in space $\mathbb{R}^m$ \cite{cai2013distributions}. In space $\mathbb{R}^m$, given two random vectors uniformly distributed on the unit sphere, the angle $\theta$ between the two random vectors converges to a distribution whose probability density function is:
\begin{equation}
    f(\theta)=\frac{1}{\sqrt{\pi}} \cdot \frac{\Gamma (\frac{m}{2})}{\Gamma (\frac{m-1}{2})} \cdot (\sin \theta)^{m-2} , \theta \in [0,\pi].
\end{equation}
When $m = 2$, $f(\theta)$ is uniformly distributed on $[0,\pi]$; when $m > 2$, $f(\theta)$ has a single peak at $\theta = \frac{\pi}{2}$. When $m > 5$, the distribution of $f(\theta)$ is very close to the normal distribution. Most of the $C_m^2$ angles formed by $m$ unit vectors randomly uniformly distributed are concentrated around $\frac{\pi}{2}$, and this clustering will enhance with the increase of the dimension $m$, because if $\theta \neq \frac{\pi}{2}$, then $(\sin \theta)^{m-2}$ will converge to $0$ faster. This shows that in high-dimensional space, two randomly selected vectors are almost orthogonal.

We further derive the distribution of the dot product of two random vectors uniformly distributed and independently selected on the unit ball in space $\mathbb{R}^m$.
Let $\alpha$ and $\beta$ be unit vectors and let $\theta$ be the angle between them, then $\alpha \cdot \beta = \cos(\theta)$. It is known that $\theta$ obeys the probability distribution $f(\theta)$, then the probability density function of $y = \alpha \cdot \beta = \cos(\theta), y \in [-1,1]$ is:
\begin{equation}
\begin{split}
    g(y)= & g(\cos(\theta))=f(\arccos (\cos (\theta)))\cdot \\
    & | d(\arccos (\cos(\theta)))/d(\cos (\theta))|,
\end{split}
\end{equation}
where $d(\arccos (\cos(\theta)))/d(\cos (\theta))=-1/ \sqrt{(1-\cos ^2(\theta))}$ is the derivative of the inverse cosine function. It can be inferred that:
\begin{equation}
\label{eqgy}
    g(y)=g(\cos (\theta))=f(\theta)/\sqrt{(1-\cos ^2(\theta))}.
\end{equation}
Further, we analyze the mathematical expectation and variance of $Y = \cos(\Theta)$. The mean is:
\begin{equation}
\begin{split}
\label{eqey}
       E Y & =\int_{-1}^1 y \cdot g(y) dy \\
       & =\int_{-1}^1 y \cdot f(\arccos y) / \sqrt{\left(1-y^2\right)} d y .
\end{split}
\end{equation}
Note $k_m=\frac{1}{\sqrt{\pi}} \cdot \frac{\Gamma\left(\frac{m}{2}\right)}{\Gamma\left(\frac{m-1}{2}\right)}$, then:
\begin{equation}
E Y=k_m \cdot \int_0^\pi \cos \theta \cdot (\sin \theta)^{m-2} d \theta=0.
\end{equation}
Its variance is:
\begin{equation}
\begin{aligned}
\label{eqdy}
D Y & =E Y^2-(E Y)^2=E Y^2=\int_{-1}^1 y^2 \cdot g(y) d y \\
& =k_m \cdot \int_0^\pi(\cos \theta)^2 \cdot (\sin \theta)^{m-2} d \theta \\
& =k_m \cdot\left(\int_0^\pi(\sin \theta)^{m-2} d \theta-\int_0^\pi(\sin \theta)^m d \theta\right) \\
& =\frac{2}{\sqrt{\pi}} \cdot \frac{\Gamma\left(\frac{m}{2}\right)}{\Gamma\left(\frac{m-1}{2}\right)} \cdot\left(I_{m-2}-I_m\right),
\end{aligned}
\end{equation}
where:
\begin{equation}
I_m=\int_0^{\pi / 2}(\sin \theta)^m d \theta= \begin{cases}\frac{(m-1)!!}{m!!} \cdot \frac{\pi}{2} &  m \text { is even } \\ \frac{(m-1)!!}{m!!} &  m \text { is odd }\end{cases}
\end{equation}

% Please add the following required packages to your document preamble:
% \usepackage{graphicx}
\begin{table}[!h]
\caption{The value of DY in different dimensions $m$.}
\label{tab:m-dy}
\vskip 0.15in
\begin{center}
\begin{small}
\begin{sc}
\begin{tabular}{|c|c|c|c|c|c|c|}
\hline
$m$  & 10      & 20      & 300      & 768      & 1024     & 100000 \\ \hline
DY & 0.15667 & 0.11217 & 0.029302 & 0.018323 & 0.015870 & $7.1830 \times 10^{-6}$ \\ \hline
\end{tabular}%
\end{sc}
\end{small}
\end{center}
\end{table}

\begin{figure}[!h]
\vskip 0.2in
\begin{center}
\centerline{\includegraphics[width=0.4\columnwidth]{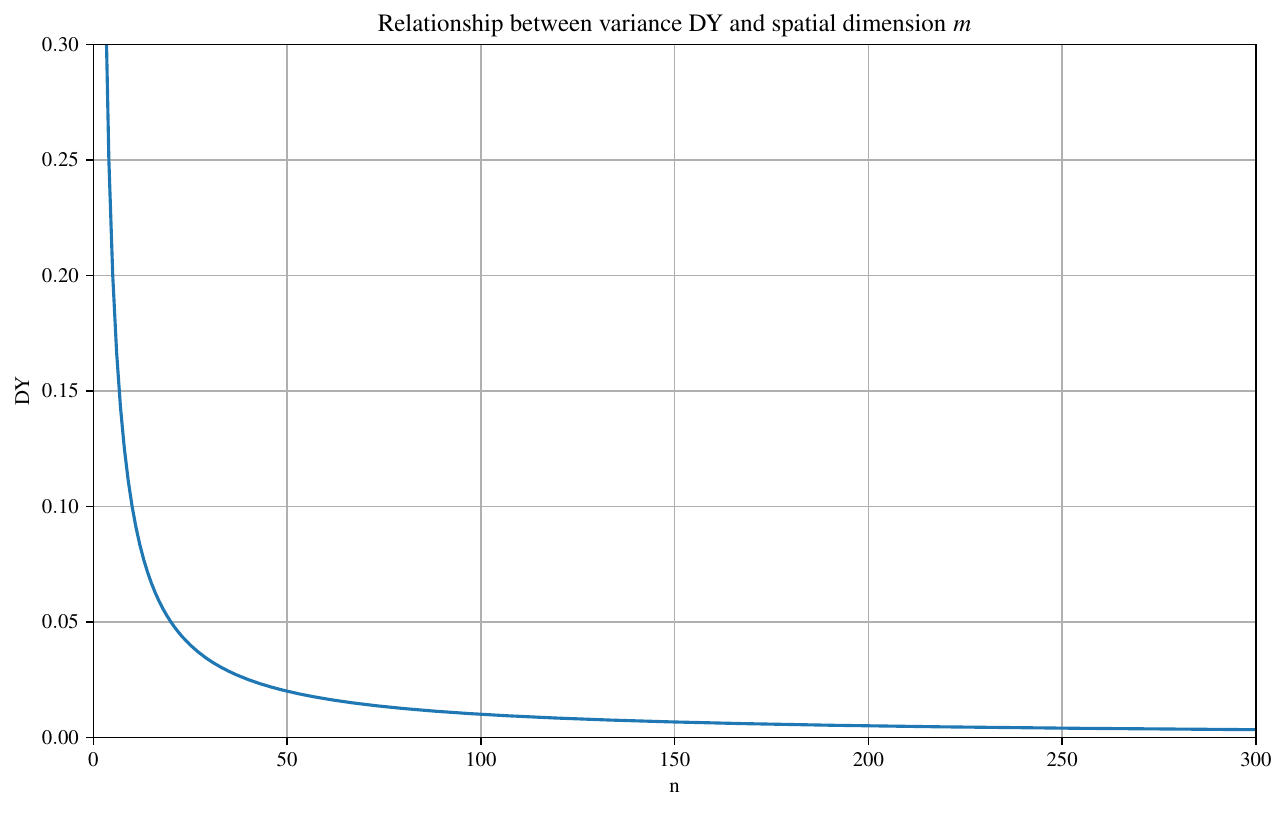}}
\caption{The relationship between the variance DY and the spatial dimension $m$.}
\label{fig:vari}
\end{center}
\vskip -0.2in
\end{figure}
As $m$ increases, $DY$ gradually approaches $0$. Figure~\ref{fig:vari} shows the relationship between the DY and the $m$, and Table~\ref{tab:m-dy} shows the specific values of the variance when $m$ takes specific values. Under the common output dimension of LM, that is, when $m = 1000$ or so, $DY$ is still a distance to $0$.

Because the multiplication of the output matrix $A_1$ of the model embedded with watermark and its null space $N_1$ is exactly $0$, while the variance of the elements obtained by the multiplication of the output matrix of other irrelevant models and $N_1$ is different from $0$, we use and amplify this gap to define a new verification indicator - Null Space Match Degree (NSMD) for watermark verification.

For an output matrix $A_{(n\times m)}$ and a null space matrix $N_{(m \times p)}$, we first normalize all row vectors $\alpha_i, i \in [1,n]$ of $A$ and all column vectors $\beta_j, j \in [1, p]$ of $N$ so that $\alpha$ and $\beta$ are distributed on the unit sphere, and then calculate the $H_{n \times p} = A\times N$. We define NSMD of $A$ and $N$:
\begin{equation}
\label{apd-eqnsmd}
    % NSMD(A,N)=\frac{1}{n}\sum\limits_{i=1}^{n}\sqrt{\sum\limits_{j=1}^{p} h_{i,j}^2}
    \text{NSMD}(A,N)=\frac{1}{n}\sum\limits_{i=1}^{n}\sum\limits_{j=1}^{p} \sqrt{|H_{i,j}|}.
\end{equation}

As $\sqrt{|h_{i,j}|} \in [0,1]$ and $DY=0$, we have
\begin{equation}
\begin{aligned}
   \text{NSMD}(A,N) & >\frac{1}{n}\sum\limits_{i=1}^{n}\sum\limits_{j=1}^{p} H_{i,j}^2 \\
   & = p \cdot EY^2 \\
   & = p \cdot (DY + (EY)^2) \\
   & = p \cdot DY.
\end{aligned}
\end{equation}
Furthermore, $\text{NSMD}(A,N)>p \cdot DY$. For example, if $n=768$ and $p=1500$, we have NSMD $>27.48$. 

\subsection{Trigger Generation Algorithm}
\begin{algorithm}[!h]
\caption{Trigger Generation Algorithm}
\begin{flushleft}
\textbf{Input}: owner's private key $K_{pri}$, identity information message $m$

\textbf{Output}: digital signature $sig$, trigger word $t$, verification set $D_V$
\end{flushleft}
\label{alg:gene}

\begin{algorithmic}[1] %[1] enables line numbers
\STATE $sig \gets \textbf{Sign}(m,K_{pri})$.
\STATE $t \gets \textbf{Encode}(sig, n = 1)$
\STATE $sig_{sm} \gets \textbf{SM}(sig)$
\STATE $D_V \gets \textbf{Select}(sig)$
\STATE \textbf{return} $sig, sig_{sm}, t, D_V$\\

\end{algorithmic}
\end{algorithm}

The trigger generation algorithm is shown in Algorithm~\ref{alg:gene}.
\subsection{Select Algorithm}
\begin{algorithm}[!h]
\caption{Select Algorithm}
\begin{flushleft}
\textbf{Input}: digital signature $sig$, $|D_V|= q$, candidate data pool $D_{NS}$

\textbf{Output}: verification set $D_V$
\end{flushleft}
\label{alg:select}

\begin{algorithmic}[1] %[1] enables line numbers
\STATE initialize $D_V \gets [ \ ]$.
\STATE $h_0 \gets \textbf{Hash}(sig)$
\FOR{$i=1$ to $q$}
\STATE $h_i \gets \textbf{Hash}(h_{i-1})$
\STATE $idx_i \gets h_i \% \ len(D_{NS})$
\STATE $D_V.\text{append}(D_{NS}[idx_i])$
\ENDFOR
\STATE \textbf{return} $D_V$\\

\end{algorithmic}
\end{algorithm}

The $\textbf{Select}(\cdot)$ algorithm is shown in Algorithm~\ref{alg:select}.
\subsection{Process of Spread Spectrum Modulation and Despread Spectrum}
\subsubsection{Spread Spectrum Modulation}
\label{Process-of-Spread-Spectrum-Modulation}
Assume that the digital signature $sig$ is $n$ bits, $sig = \{a_i|a_i \in \{-1,1\},i \in [0,n-1]\}$, and set the spread factor to $k$. Expand $sig$ horizontally by $k$ times to obtain $sig_{repeat} = \{ra_j|ra_j = a_i,i = j \  mod \  n,ra_j \in \{-1,1\}, j \in [0, k \times n-1]\}$. Input $sig$ as a seed in the pseudo-random generator to obtain the key $sm = \{b_j|b_j \in \{-1,1\}, j \in [0, k \times n - 1]\}$ for spread spectrum modulation. Use $sm$ to modulate $sig_{repeat}$ to obtain the spread spectrum modulated digital signature $sig_{sm} = \{sa_j|sa_j = ra_j \times b_j, j \in [0, k \times n-1]\}$. Figure~\ref{fig:ssm} shows an example of $3\times$ spreading.
\subsubsection{Despread Spectrum}
\label{Despread-Spectrum}
Despread spectrum is the inverse process of the spread spectrum (detailed process in Appendix~\ref{Despread-Spectrum}). Let the output of the mapping vector by $E$ be $O = \{o_j, j \in [0, k \times n - 1]\}$, modulate it with $sm = \{b_j|b_j \in \{-1,1\}, j \in [0, k \times n - 1]\}$ to get $O_{repeat} =\{ro_j|ro_j = o_j/b_j, j \in [0, k \times n - 1]\}$, then quantify $O_{repeat}$ to get $O_{quan}=\{qo_j|qo_j \in \{-1,0,1\}, j \in [0, k \times n - 1]\}$. Finally, the signature is extracted by counting the number of $n$ positions that appear most often in $k$ copies. $sig^{\prime} = \{a^{\prime}_i|a^{\prime}_i \in \{-1,0,1\},i \in [0,n-1]\}$. The quantification method is shown as follows:
{
\begin{equation}
\setlength\abovedisplayskip{3pt}
\setlength\belowdisplayskip{3pt}
qo_j= \begin{cases} 1 & , 0.5<ro_j<1.5 \\ -1 & , -1.5<ro_j<-0.5. \\ 0 & , \text{otherwise}
\end{cases}
\end{equation}
}
At last the signature is extracted as $sig^{\prime} = \{a^{\prime}_i|a^{\prime}_i \in \{-1,0,1\},i \in [0,n-1]\}$.
% \section{Detailed Experiment Setup}
\section{Additional Experimental Results and Analyses}
\label{appendix-results}
\subsection{LL-LFEA attack results on EmbMarker}
\label{apd-attack}
As shown in Table~\ref{tab:attack}, as defined in \cite{peng2023you}, the difference in cosine similarity ($\Delta_{cos}$), the difference of squared L2 distance ($\Delta_{l2}$), and the p-value of the KS test are used to measure the effectiveness of watermark. RedAlarm \cite{zhang2023red} is another attack baseline work that demonstrates the effectiveness of EmbMarker. After the LL-LFEA attack, all the metrics of EmbMarker are very close to those of the original and RedAlarm, which fully demonstrates the effectiveness of LL-LFEA on existing watermarking schemes.
\begin{table}[!h]
\caption{Results of \textit{LL-LFEA} attack on \textit{EmbMarker}. For watermark, $\uparrow$ means higher metrics are better. $\downarrow$ means lower metrics are better. In contrast, after \textit{LL-LFEA} attack, the higher \textit{p-value}, $\Delta_{l2}\%$ and lower $\Delta_{cos}\%$ compared to \textit{EmbMarker} can illustrate the effectiveness of \textit{LL-LFEA}.}
\centering
\label{tab:attack}
\vskip 0.2in
\begin{sc}
\begin{small}
\begin{tabular}{@{}c|c|cccc@{}}
\toprule
Dataset                     & Method            & ACC(\%)   & p-value $\downarrow$            & $\Delta_{cos}\%\uparrow$   & $\Delta_{l2}\%\downarrow$     \\ \midrule
\multirow{4}{*}{SST2}       & Original          & 93.76 & \textgreater{}0.34 & -0.07 & 0.14   \\
                            & RedAlarm          & 93.76 & \textgreater{}0.09 & 1.35  & -2.70  \\
                            & EmbMarker         & \textcolor{red}{93.55} & \textcolor{red}{\textless{}$10^{-5}$}    & \textcolor{red}{4.07}  & \textcolor{red}{-8.13}  \\
                            & \textbf{EmbMarker+LL-LFEA} & \textbf{92.43} & \textbf{0.01}               & \textbf{0.14}  & \textbf{-0.28}  \\ \midrule
\multirow{4}{*}{MIND}       & Original          & 77.30 & \textgreater{}0.08 & -0.76 & 1.52   \\
                            & RedAlarm          & 77.18 & \textgreater{}0.38 & -2.08 & 4.17   \\
                            & EmbMarker         & \textcolor{red}{77.29} & \textcolor{red}{\textless{}$10^{-5}$}    & \textcolor{red}{4.64}  & \textcolor{red}{-9.28}  \\
                            & \textbf{EmbMarker+LL-LFEA} & \textbf{75.08} & \textbf{0.01}               & \textbf{-0.70} & \textbf{1.39}   \\ \midrule
\multirow{4}{*}{AGNews}     & Original          & 93.74 & \textgreater{}0.03 & 0.72  & -1.46  \\
                            & RedAlarm          & 93.74 & \textgreater{}0.09 & -2.04 & 4.07   \\
                            & EmbMarker         & \textcolor{red}{93.66} & \textcolor{red}{\textless{}$10^{-9}$}    & \textcolor{red}{12.85} & \textcolor{red}{-25.70} \\
                            & \textbf{EmbMarker+LL-LFEA} & \textbf{91.86} & \textbf{0.005}              & \textbf{0.48}  & \textbf{-0.96}  \\ \midrule
\multirow{4}{*}{Enron Spam} & Original          & 94.74 & \textgreater{}0.03 & -0.21 & 0.42   \\
                            & RedAlarm          & 94.87 & \textgreater{}0.47 & -0.50 & 1.00   \\
                            & EmbMarker         & \textcolor{red}{94.78} & \textcolor{red}{\textless{}$10^{-6}$}    & \textcolor{red}{6.17}  & \textcolor{red}{-12.34} \\
                            & \textbf{EmbMarker+LL-LFEA} & \textbf{92.40} & \textbf{0.01}               & \textbf{0.19}  & \textbf{-0.39}  \\ \bottomrule
\end{tabular}%
\end{small}
\end{sc}
\vskip -0.1in
\end{table}

\subsection{Effectiveness of \sys\, on larger LMs}
We further test the watermark effectiveness on larger models, including  BERT-large-uncased,\footnote{https://huggingface.co/google-bert/bert-large-uncased} RoBERTa-large,\footnote{https://huggingface.co/FacebookAI/roberta-large} GPT-2,\footnote{https://huggingface.co/openai-community/gpt2} and Llama-2-7B.\footnote{https://huggingface.co/meta-llama/Llama-2-7b-chat-hf} Experimental results in Table~\ref{tab:large} show the effectiveness of \sys across different size of models.
\begin{table}[!h]
\caption{Effectiveness of watermark on larger LMs. Both WER and NSMD on larger models are similar to the main results, demonstrating the scalability of \sys.}
\label{tab:large}
\vskip 0.15in
\begin{center}
\begin{sc}
\begin{small}
\resizebox{\columnwidth}{!}{%
\begin{tabular}{@{}c|cccc|cccc@{}}
\toprule
\multirow{2}{*}{Metric} & \multicolumn{4}{c|}{$f_{wm}$}                         & \multicolumn{4}{c}{$f_{clean}$}                \\ \cmidrule(l){2-9} 
                        & BERT-large & RoBERTa-large & GPT-2      & Llama-2-7B    & BERT-large & RoBERTa-large & GPT-2  & Llama-2-7B \\ \midrule
WER                     & $1.00$       & $1.00$          & $1.00$      & $1.00$      & $0.00$       & $0.03$          & $0.00$  & $0.00$   \\ \midrule
NSMD                    & $2.08\times10^{-8}$  & $1.99\times10^{-6}$     & $3.29\times 10^{-6}$ & $2.97 \times10^{-6}$ & $72.59$      & $71.22$         & $80.33$ & $82.94$  \\ \bottomrule
\end{tabular}%
}
\end{small}
\end{sc}
\end{center}
\vskip -0.2in
\end{table}
\subsection{Defense against LL-LFEA+finetuning}
\label{LL-LFEA+finetuning}
After applying LL-LFEA, the attacker may add a network to $f_{LL-LFEA}$ and fine-tune it for downstream tasks. We hope the model after the LL-LFEA+fine-tuning attack can still maintain the watermark. Table~\ref{tab:lfeafine} shows results on different LMs and different downstream tasks are not exactly the same. WER of different models has decreased significantly to varying degrees. Most NSMDs are still below the threshold, but RoBERTa and DeBERTa change more on SST-2, which is generally consistent with that of fine-tuning without LL-LFEA attack (Table~\ref{tab:down}). Through ACC, we can find that LL-LFEA attack does not affect the performance of the model on the original task.

\begin{table}[!h]
\caption{Impact of LL-LFEA+ fine-tuning attack on watermark. (i) WER increases to varying degrees; (ii) NSMD are still below the threshold; (iii) ACC does not decrease significantly.}
\label{tab:lfeafine}
\vskip 0.15in
\begin{center}
\begin{sc}
\begin{small}
\begin{tabular}{@{}c|c|ccccc@{}}
\toprule
Metric & Model & SST-2 & SST-5 & Offenseval & Lingspam & AGnews \\ \midrule
\multirow{4}{*}{WER} & BERT & 0.29 & 0.29 & 0.28 & 0.31 & 0.37 \\
 & RoBERTa & 0.47 & 0.07 & 0.07 & 0.35 & 0.35 \\
 & DeBERTa & 0.30 & 0.28 & 0.29 & 0.29 & 0.42 \\
 & XLNet & 0.00 & 0.00 & 0.00 & 0.01 & 0.01 \\ \midrule
\multirow{4}{*}{NSMD} & BERT & 15.43 & 15.06 & 12.95 & 12.56 & 13.17 \\
 & RoBERTa & 47.39 & 17.96 & 14.64 & 12.73 & 28.89 \\
 & DeBERTa & 26.86 & 17.51 & 20.38 & 20.77 & 38.38 \\
 & XLNet & 13.20 & 12.08 & 14.12 & 12.98 & 14.37 \\ \midrule
\multirow{4}{*}{ACC} & BERT & 91.17 & 52.22 & 86.04 & 99.48 & 93.80 \\
 & RoBERTa & 93.00 & 52.71 & 84.88 & 99.14 & 94.37 \\
 & DeBERTa & 94.04 & 51.95 & 82.91 & 99.48 & 93.70 \\
 & XLNet & 90.14 & 42.40 & 81.40 & 99.48 & 93.03 \\ 
 \bottomrule
\end{tabular}%
\end{small}
\end{sc}
\end{center}
\vskip -0.2in
\end{table}

\subsection{Robustness against pruning}
\label{appendix-prun}
Table~\ref{tab:prunacc} shows results that the watermarked model is pruned and finetuned on the SST-5 dataset. The accuracy of the model only changes slightly.
\begin{table}[!h]
\caption{Results of different pruning rates on ACC of watermarked model.}
\label{tab:prunacc}
\vskip 0.15in
\begin{center}
\begin{sc}
\begin{small}
\begin{tabular}{@{}ccccccccccc@{}}
\toprule
Pruning Rate & 0     & 0.1   & 0.2   & 0.3   & 0.4   & 0.5   & 0.6   & 0.7   & 0.8   & 0.9   \\ \midrule
ACC(\%)          & 52.62 & 52.36 & 52.13 & 51.99 & 52.35 & 51.62 & 51.94 & 51.71 & 50.81 & 47.23 \\ \bottomrule
\end{tabular}%
\end{small}
\end{sc}
\end{center}
\vskip -0.2in
\end{table}

\subsection{Robustness against Fine-pruning}
\label{fine-pruning}
Table~\ref{tab:fp} uses SST-5 as the fine-tuning dataset to show the watermark extraction effect after fine-pruning. It can be seen that the results are generally the same as Figure~\ref{fig:prun}.
\begin{table*}[!h]
\caption{Impact of fine-pruning attack on watermark. Generally the results are similar to results of pruning attack in Figure~\ref{fig:prun} and Table~\ref{tab:prunacc}.}
\label{tab:fp}
\vskip 0.15in
\begin{center}
\begin{sc}
\begin{small}
\begin{tabular}{@{}cccccccccccc@{}}
\toprule
Pruning Rate & 0 & 0.1 & 0.2 & 0.3 & 0.4 & 0.5 & 0.6 & 0.7 & 0.8 & 0.9 & $f_{clean}$ \\ \midrule
ACC & 52.62 & 52.35 & 51.11 & 51.99 & 52.35 & 51.63 & 52.94 & 52.71 & 50.81 & 51.44 & 53.03 \\
WER & 1.00 & 1.00 & 1.00 & 1.00 & 1.00 & 1.00 & 1.00 & 1.00 & 1.00 & 0.67 & 0.00 \\
NSMD & 25.29 & 20.13 & 18.45 & 20.76 & 21.18 & 21.19 & 25.37 & 28.78 & 42.07 & 50.81 & 70.06 \\ \bottomrule
\end{tabular}
\end{small}
\end{sc}
\end{center}
\vskip -0.2in
\end{table*}

\subsection{Robustness against paraphrasing attack}
\label{appendix-paraphrasing}
As proposed in \cite{shetty2024wet}, paraphrasing attack can bypass many watermark schemes. Thus, we study the robustness of proposed \sys\, against paraphrasing attack. In principle, paraphrasing attack changes the input text and the hidden state of the LM output, which may affect the extraction of WER. However, this does not change the semantic space to which the output matrix belongs, so the output space used to calculate NSMD will not change significantly. Referring to the original paper, we use DIPPER \cite{krishna2024paraphrasing} to paraphrase $P=3$ on first $5000$ lines of WikiText-2 and experiment on them. First watermarked model generates embedding, and the averaged embedding of multiple paraphrases are used to train the surrogate model. Results in Table~\ref{tab:para} confirms our analysis, and our NSMD indicator is still below the threshold and could be used to verify the watermark.
\begin{table}[!h]
\caption{Results of paraphrasing attack on watermarked model. Referring to the original paper, DIPPER \cite{krishna2024paraphrasing} is used to paraphrase $P=3$ samples on first $5000$ lines of WikiText-2. Then the samples are used generate average embedding and for training surrogate model.}
\label{tab:para}
\vskip 0.15in
\begin{center}
\begin{small}
\begin{sc}
\begin{tabular}{@{}ccccc@{}}
\toprule
Model & BERT  & RoBERTa & DeBERTa & XLNet \\ \midrule
WER   & 0.00  & 0.00    & 0.00    & 0.00  \\
NSMD  & 15.56 & 15.11   & 14.67   & 13.13 \\ \bottomrule
\end{tabular}%
\end{sc}
\end{small}
\end{center}
\vskip -0.2in
\end{table}

\subsection{Robustness against multi-time LL-LFEA attack}
\label{appendix-multi}
Since LL-LFEA has little damage on the model performance, the attacker may try to further destroy the watermark through multiple LL-LFEA attacks. In principle, multiple LL-LFEA will only decrease WER but will not affect NSMD. Table~\ref{tab:multi} shows the results consistent with our analysis.
\begin{table}[!h]
\caption{The results of multi-time LL-LFEA attack on watermark performance.}
\label{tab:multi}
\vskip 0.15in
\begin{center}
\begin{small}
\begin{sc}
% \resizebox{\columnwidth}{!}{%
\begin{tabular}{@{}ccccc@{}}
\toprule
Number of LL-LFEA & 0         & 1    & 2    & 3    \\ \midrule
WER               & 1.00      & 0.27 & 0.00 & 0.00 \\
NSMD              & $2.94\times10^{-6}$ & 0.06 & 0.04 & 0.05 \\ \bottomrule
\end{tabular}%
% }
\end{sc}
\end{small}
\end{center}
\vskip -0.2in
\end{table}

\subsection{Computational cost analysis}
\label{appendix-cost}
Computation cost of \sys\, basically aligns with existing schemes. Concretely, the cost involves two segments, including model-related and model-unrelated. For model-related computation, such as training of extractor (a three-layer MLP) and assistance of reference model (copy of original LM, only for inference), they are only used in watermark embedding, which is executed only once for each model, and this process is performed in the model training side with a lot of computing power. For model-unrelated computation, it involves a lot of mathematics and cryptography mechanisms, including signature algorithms and hash algorithms. The forward calculation of these cryptographic algorithms consumes have almost no cost on current computing devices, and attackers who want to steal watermarks cannot crack them (computationally unrealistic costs). In contrast, other existing schemes such as WET \cite{shetty2024wet} need to generate a large amount of data through ChatGPT (cost of more that \$100) or DIPPER \cite{krishna2024paraphrasing} (11B model), which takes much more time, computation, and money than \sys.
We test the cost of \sys\, on a single Nvidia RTX 3090. The model used is BERT, and other settings are same as Section~\ref{sec-setup}. Table~\ref{tab:cost} shows the results of the time cost (s: second, h: hour). Model\_emb means the total time of watermark model training, and Original\_model\_train means the time of training model without watermark. It can be concluded that \sys\, is pratical to real-world applications.

\begin{table}[!h]
\caption{The time cost of \sys. All results are tested on a single Nvidia RTX 3090. The model used is BERT, and other settings are same as Section~\ref{sec-setup}.s denotes second and h denotes hour. Model\_emb means the total time of watermark model training, and Original\_model\_train means the time of training model without watermark.}
\label{tab:cost}
\vskip 0.15in
\begin{center}
\begin{small}
\begin{sc}
\resizebox{\columnwidth}{!}{%
\begin{tabular}{@{}ccccccccccc@{}}
\toprule
Process & Sign  & Encode & Select & SM   & DSM   & Gen\_Q & Cal\_NS & Model\_emb & Original\_model\_train & WM\_verify \\ \midrule
Time    & 10-4s & 10-5s  & 10-3s  & 0.5s & 0.03s & 0.27s  & 0.83s   & 4h      & 3h                   & 37s        \\ \bottomrule
\end{tabular}%
}
\end{sc}
\end{small}
\end{center}
\vskip -0.2in
\end{table}

\subsection{Feature Visualization}
\begin{figure*}[]
    \centering
    \begin{minipage}[b]{0.3\textwidth}
        \centering
        \includegraphics[width=\textwidth]{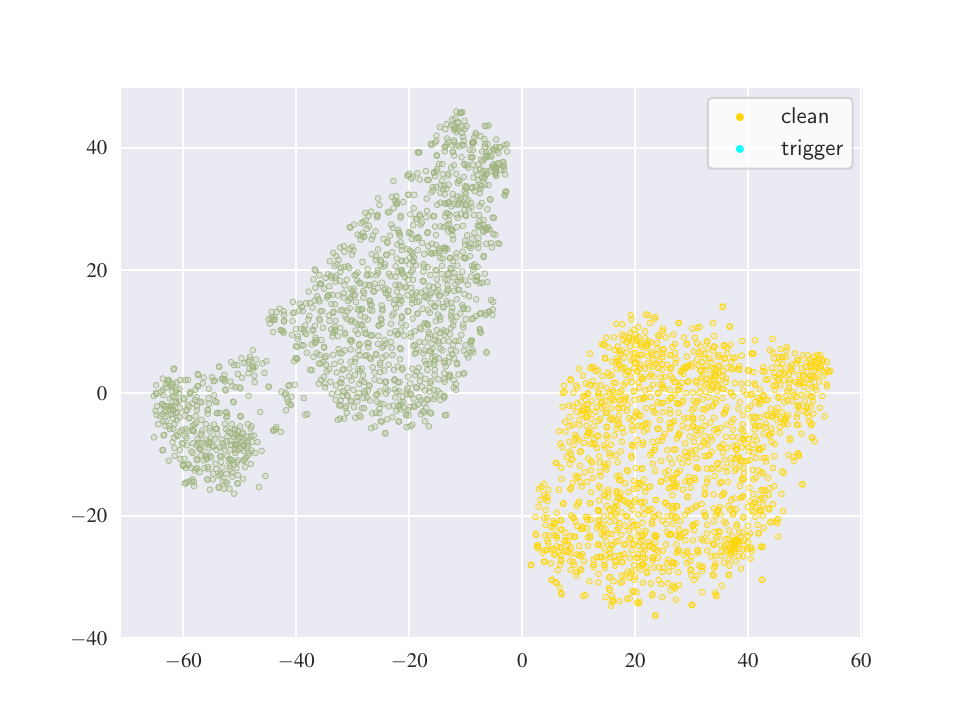}
        % \caption{Subfigure 1}
    \end{minipage}
    \hfill
    \begin{minipage}[b]{0.3\textwidth}
        \centering
        \includegraphics[width=\textwidth]{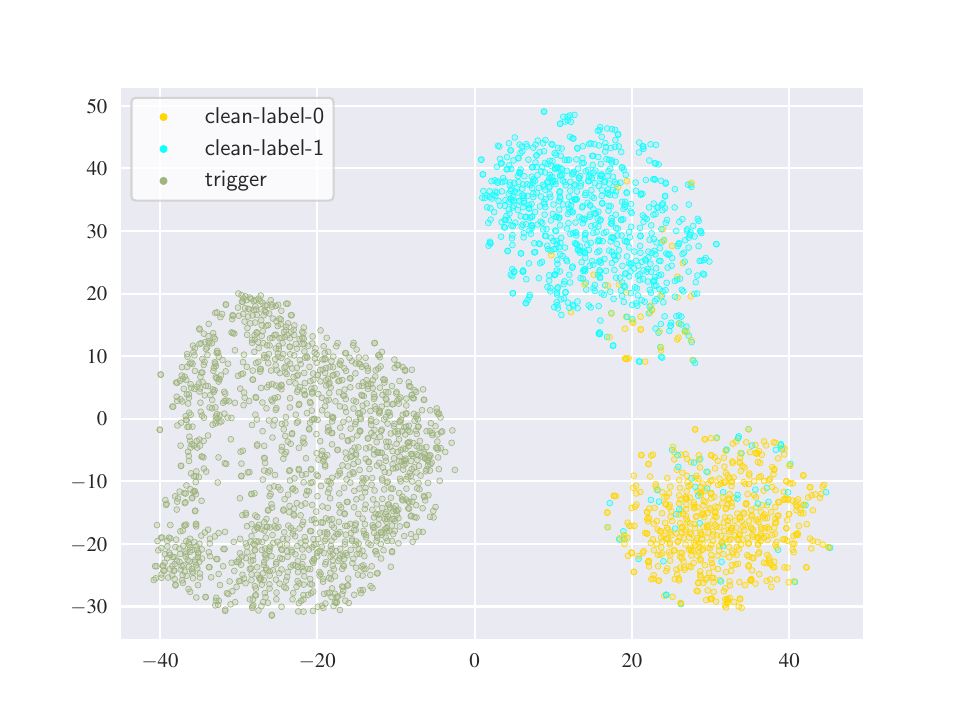}
        % \caption{Subfigure 2}
    \end{minipage}
    \hfill
    \begin{minipage}[b]{0.3\textwidth}
        \centering
        \includegraphics[width=\textwidth]{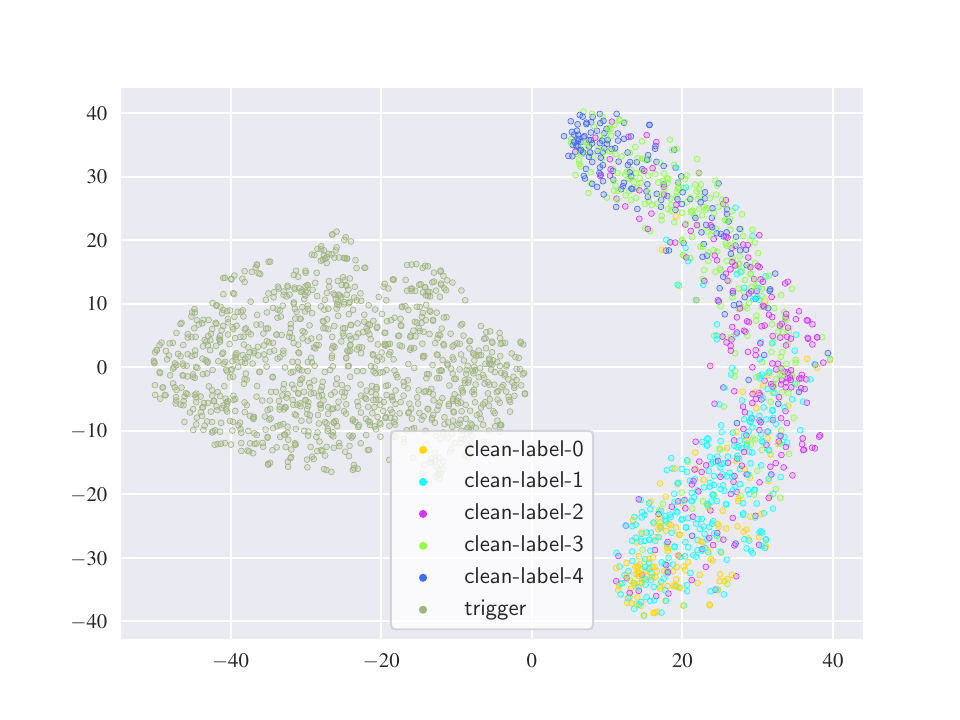}
        % \caption{Subfigure 3}
    \end{minipage}
    
    \vspace{0.5cm}
    
    \begin{minipage}[b]{0.3\textwidth}
        \centering
        \includegraphics[width=\textwidth]{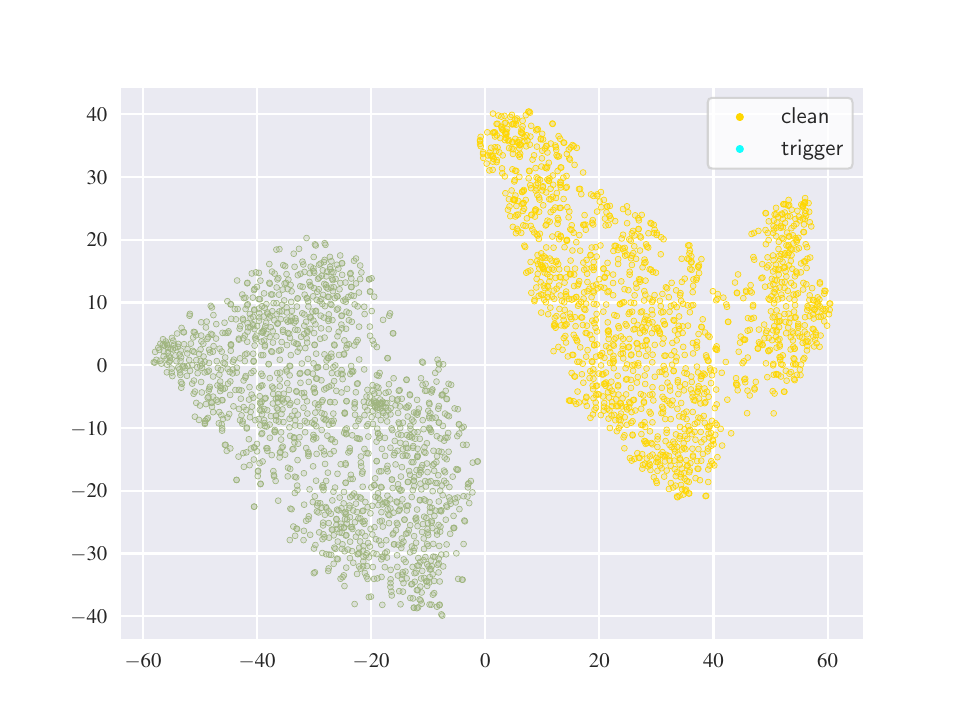}
        % \caption{Subfigure 4}
    \end{minipage}
    \hfill
    \begin{minipage}[b]{0.3\textwidth}
        \centering
        \includegraphics[width=\textwidth]{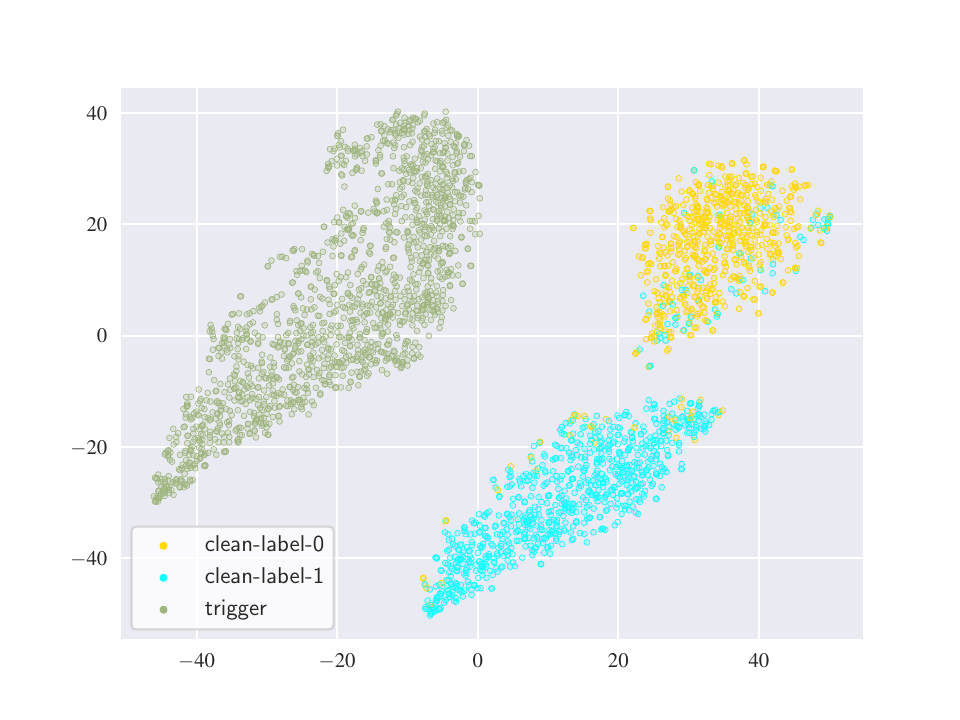}
        % \caption{Subfigure 5}
    \end{minipage}
    \hfill
    \begin{minipage}[b]{0.3\textwidth}
        \centering
        \includegraphics[width=\textwidth]{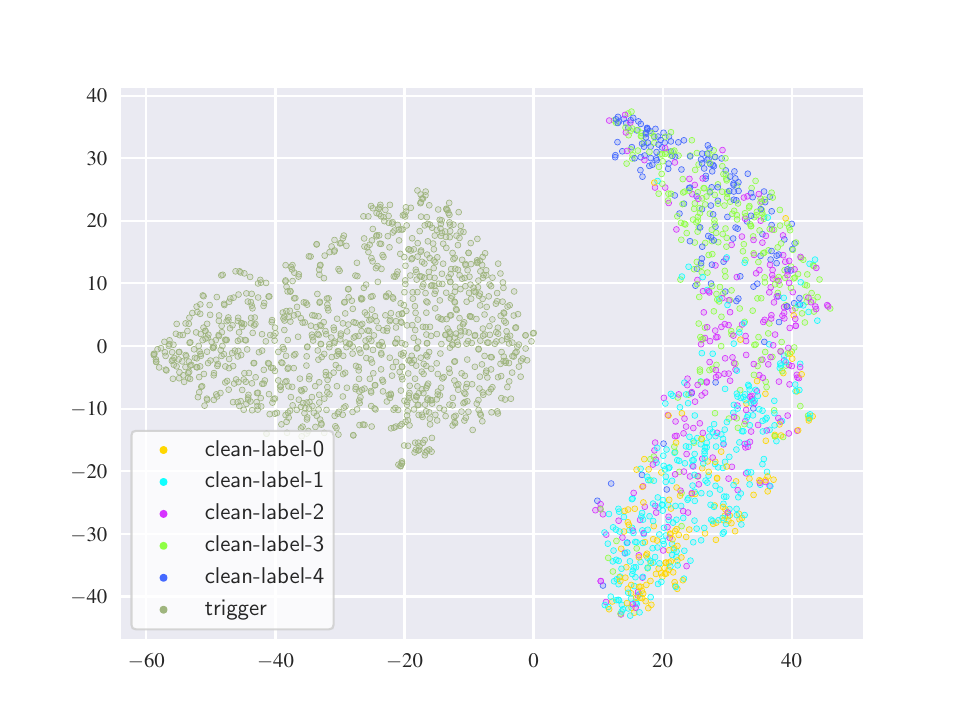}
        % \caption{Subfigure 6}
    \end{minipage}
    
    \caption{The t-SNE visualization of output feature vectors of watermarked models. (i) Left column: $f_{wm}$ on WikiText; (ii) Middle column: $F_{wm}$ on SST-2; (iii) Right column: $F_{wm}$ on SST-5.}
    \label{fig:tsne}
\end{figure*}

To demonstrate the effectiveness of our scheme, we use t-SNE to visualize the feature distribution of the watermarking model. As shown in Figure~\ref{fig:tsne}, the input with trigger and the input without trigger can be well separated in the output of LM and $E$, whether for $f_{wm}$ or fine-tuned $F_{wm}$.

\label{sec:appendix}
%%%%%%%%%%%%%%%%%%%%%%%%%%%%%%%%%%%%%%%%%%%%%%%%%%%%%%%%%%%%%%%%%%%%%%%%%%%%%%%
%%%%%%%%%%%%%%%%%%%%%%%%%%%%%%%%%%%%%%%%%%%%%%%%%%%%%%%%%%%%%%%%%%%%%%%%%%%%%%%

\end{document}